\documentclass[sn-basic]{sn-jnl}

\usepackage{graphicx}%
\usepackage{multirow}%
\usepackage{amsmath,amssymb,amsfonts}%
\usepackage{amsthm}%
\usepackage[title]{appendix}%
\usepackage{xcolor}%
\usepackage{textcomp}%
\usepackage{manyfoot}%
\usepackage{booktabs}%
\usepackage{algorithm}%
\usepackage{algorithmicx}%
\usepackage{algpseudocode}%
\usepackage{listings}%
\usepackage{mathtools}
\usepackage{bbm}
\usepackage{tikz}
\usetikzlibrary{arrows.meta, positioning}
\usepackage{subcaption}

\theoremstyle{thmstyleone}%
\theoremstyle{thmstyletwo}%

\theoremstyle{thmstylethree}%
\newcommand{\pkg}[1]{\textbf{#1}}        
\newcommand{\code}[1]{\texttt{#1}}
\begin{document}

\title[Analysis of Multifractional Stochastic Processes with Rmfrac]{Simulation and Analysis of Multifractional Stochastic Processes with R Package Rmfrac}

\author*[1]{\fnm{Andriy} \sur{Olenko} \orcid{https://orcid.org/0000-0002-0917-7000}} \email{a.olenko@latrobe.edu.au}

\author[1]{\fnm{Nemini} \sur{Samarakoon} \orcid{https://orcid.org/0009-0000-2107-1973}} \email{n.wijesinghesamarakoon@latrobe.edu.au}
\equalcont{These authors contributed equally to this work.}

\affil*[1]{\orgdiv{Department of Mathematical and Physical Sciences}, \orgname{La Trobe University}, \orgaddress{\city{Melbourne}, \country{Australia}}}

\abstract{Brownian motion and fractional Brownian motion have been widely applied in statistical modeling in finance, telecommunication, network traffic, neuroscience, physics, and other fields. More realistic models for real time series data, such as multifractional processes, generalize these classical models by allowing their regularity to vary over time. A new class of  Gaussian Haar-based multifractional processes, which utilizes the Haar wavelet series representation, was recently introduced. It significantly extends the range of available models by incorporating more general classes of Hurst functions. The Rmfrac package was developed to simulate multifractional time series. The package also comprises several functions for the analysis and visualization of time series. It includes the estimation of the Hurst function and local fractal dimension, clustering realizations and computing various geometric statistics of these time series. The package also offers a Shiny application to visualize simulation and estimation results. The article presents an overview of the Rmfrac package and exemplifies its main functionalities.}




\maketitle

\section{Introduction}\label{sec1}

In the following, we interchangeably use the terms multifractional processes and time series. In most cases, the former refers to observations at arbitrary time moments, while the latter refers to observations on a regular time grid.

Brownian motion (Bm) is one of the main stochastic models, especially in the study of functional time series, with wide-ranging applications. The mathematical model for Bm was proposed by~\cite{wiener1923differential}. It is frequently used in finance to model stock and derivative securities market prices~\citep{davis2006louis,meyer2003strategic}, has numerous applications in biology, physics, and engineering~\citep{Bier01111997, saffman1975brownian, vedavathi2022chemical}, just to name a few. 

The fractional Brownian motion (fBm) proposed by~\cite{mandelbrot1968fractional} extends the idea of Bm, in particular, by introducing the long memory property and specifying roughness of realizations through a constant Hurst parameter. These properties allow fBm to model persistence in time series and provide a more accurate representation of real-world data, with applications ranging from financial market forecasting and trading strategies \citep{Garcin03082022}, stock and option pricing \citep{ItofBmDai}, queuing theory and network traffic  \citep{decreusefond1998fractional, 400651}, hydrology \citep{MolzfBm}, and beyond.

Multifractional processes have been proposed as a generalization for fBm, allowing the Hurst parameter to change with time and control the regularity and long range dependence. Due to this flexibility, multifractional processes have demonstrated empirical effectiveness in capturing market volatility and modeling option and stock prices \citep{bianchi2014asset,  mattera2021option, araneda2024multifractional, peng2018general}. They also found numerous applications in spatial and environmental modeling \citep{ lee2013characterization, echelard2010terrain, levy2013beyond, mastalerz2018application}. For examples of other applications, see \cite{bianchi2012modeling}, \cite{gaci2011two}, \cite{song2020multifractional} and \cite{Kang}.

The construction of multifractional processes ranges from models based on integral approximations \citep{peltier1995multifractional, benassi1997elliptic, muniandy2001modeling, lacaux2004real, ayache2018new} to functional random series representations  \citep{ayache2005multifractional, ayache2007wavelet, guevel2012ferguson}. Most of these constructions are very difficult to use in practical applications, simulations and statistical analysis. The recently proposed model of Gaussian Haar-based multifractional processes \citep{ayache2025} involves a new construction based on the Haar wavelet approach. It offers a simple framework for efficient simulations and analysis, and can provide a broad class of Hurst functions.

Some R packages currently available on CRAN include functions that support the simulation of classical fractional models (Bm, fBm, Bm bridges, and fractional Gaussian noise), see \pkg{Sim.DiffProc} \citep{simdiffproc}, \pkg{sde} \citep{sde}, \pkg{longmemo} \citep{longmemo}, \pkg{simts} \citep{simts}, \pkg{e1071} \citep{e1071}. 

However, only a few software tools have been developed for simulating and analyzing multifractional processes and time series in different programming environments. In MATLAB, there is a function for simulating Riemann–Liouville multifractional Brownian motion (mBm)  \citep{giannit2024mbm}. The toolbox \pkg{Fraclab} \citep{fraclab-software} developed for signal and image processing, which runs via the MATLAB interface, includes a function for mBm simulations. This provides an approximation to the mBm. However, beyond a few hundred points, the simulations are increasingly time-consuming.

The Python package \pkg{stochastic} \citep{flynnmbm} provides a function for mBm simulations, implementing the method described by \cite{muniandy2001modeling}, which approximates the Riemann–Liouville integral via discretization. The Python package \pkg{fractal-analysis} \citep{yujiadingpkg} provides two functions for simulating mBm. The first uses the Wood and Chan circulant matrix method with kriging~\citep{ChanSimMBM}, while the second applies the Lamperti transformation with kriging. These approaches are approximate due to discretization of the Hurst range and kriging interpolation, with reduced accuracy near boundaries or when the true Hurst parameter lies outside the simulated range.
 
The R package \pkg{fractalRegression} \citep{FractalReg}(currently removed from CRAN) includes a function to simulate mBm. It uses an approximate method for mBm simulation, and the sample paths of the realizations do not exhibit the roughness accurately. The R~package \pkg{FracSim} (currently archived on CRAN, not updated since 2009, and archived versions have installation issues) includes functions to simulate the non-Gaussian multifractional model by~\cite{JSSv014i18}.

To the best of our knowledge and following an extensive search, there is no software that accurately estimates the roughness of realizations and implements it in other statistical analyses related to Hurst functions. The R packages currently available on CRAN that provide Hurst estimates are \pkg{pracma} \citep{pracma} and \pkg{PerformanceAnalytics} \citep{PerformanceAna}. These packages only provide estimates of a single constant Hurst value. Moreover, there are GitHub issues reporting discrepancies in the results of the function in \pkg{PerformanceAnalytics}.

In general, there is a lack of software that supports the complete workflow for multifractional processes, including accurate simulations for a wide range of Hurst functions, precise estimation of the Hurst function, and incorporating it in other related statistical tasks.

The \pkg{Rmfrac} \citep{rmfrac} package addresses the aforementioned problems. First, it provides a comprehensive toolkit that includes the simulation of realizations of various fractional and multifractional time series, estimation of the Hurst function and fractal dimensions of time series, and corresponding visualizations. \pkg{Rmfrac} supports accurate multifractional simulation for a wide class of Hurst functions, including irregular and discontinuous ones. It provides estimates of time-varying roughness for either simulated processes or user-provided time series. Secondly, for non-stationary realizations, it computes theoretical covariance functions for given Hurst functions or their estimators using sampled time series realizations. Further statistical analysis includes clustering time series based on the estimated Hurst functions and estimating various geometric statistics of the realizations. 

Compared to the mentioned existing software implementations, which require lengthy times for simulations and multifractionality estimation, \pkg{Rmfrac} employs an efficient computational approach, producing results within a reasonable time frame. \pkg{Rmfrac} also provides a Shiny app that visualizes simulations, estimations, and main statistical functions for both simulated processes and user-provided time series.

The article is organized as follows. Section~\ref{sec2} provides the theoretical concepts of standard and fractional Brownian motions and the Gaussian Haar-based multifractional processes. It introduces the Hurst function, local fractal dimension estimation, and the theory behind the geometric statistics implemented in \pkg{Rmfrac}. Section~\ref{sec3} discusses the structure of the \pkg{Rmfrac} package and its core functions with examples. Section~\ref{sec4} concludes the article.

\section{Theoretical concepts}\label{sec2}

This section introduces the theoretical concepts and basic stochastic and statistical models used in the functions of the \pkg{Rmfrac} package. 

\subsection{Standard and fractional Brownian motion}

Brownian motion is a continuous-time zero-mean stochastic process $\{X(t)\}_{t\geq0}$ that has stationary independent increments $X(s+t)-X(t) \sim \mathcal{N}(0,s)$. Its generalization, the fractional Brownian motion  $\{X_{H}(t)\}_{t\geq 0}$, is a Gaussian process that is centered, self-similar and has the covariance function 
$$E(X_{H}(t)X_{H}(s)) = \frac{1}{2}\left( t^{2H}+ s^{2H} -|t-s|^{2H} \right), \quad t,s \geq 0,$$
where $H \in (0,1)$ is the Hurst parameter. Compared to Bm, increments of fBm are stationary, but dependent. The constant Hurst parameter determines the regularity of the fBm. For Bm it is $H=0.5$. 

The modification of Bm is the Brownian Bridge, which starts and terminates at the specified values $X^S(0)$ and $X^S(T)=a$. It is given by
\begin{equation*}
    X^S(t) = X(t) - \dfrac{t}{T}(X(T)-a),  \quad t \in [0,T],
\end{equation*}
where $X(t)$ is Bm. Similarly, the fractional Brownian bridge on $[0,T],$ which terminates at $a,$ is defined as
\begin{equation*}
    X_H^S(t) = X_H(t) - \dfrac{1}{2}\left(X_H(T)-a\right)\left( 1+ \left(\dfrac{t}{T}\right)^{2H}-\left(1-\dfrac{t}{T}\right)^{2H}\right),
\end{equation*}
where $X_H(t)$ is fBm \citep{FBb_Binachi}.

Multifractional processes are generalizations of fBm. Their regularities can vary in time, and instead of constant Hurst parameters, the Hurst function quantifies the roughness of a process in a neighborhood of a chosen time point. 

In general, for a stochastic process $X(t)$ with continuous sample paths, the pointwise H\"older exponent at point $t$
is defined~by
\[\alpha_X(t) = \sup \left\{ \alpha \in \mathbb{R}_+ : \limsup_{h \to 0} \frac{|X(t + h) - X(t)|}{|h|^{\alpha}} = 0 \right\} 
\]
and is used to quantify the roughness of the process in a neighbourhood of any chosen fixed time point.
 
\subsection{Gaussian Haar-based multifractional processes}

The approach of utilizing Haar wavelet series in constructing multifractional processes was introduced by~\cite{ayache2025}. 

Let $H(t)$, $t \in [0,1]$, be the Hurst function. If $H(t)$ is a smooth function, we use the functions $H_{j}(\cdot)\equiv H(\cdot)$, for all $j = 0,1,\dots$ . For those $H(t)$ that have discontinuities, we use a sequence $\{H_{j}(\cdot)\}$ of continuous functions that $\displaystyle \lim_{j \to \infty} H_j(t)= H(t)$ in all points of continuity of $H(\cdot)$. In the following, we use the notation $H_{j,k} = H_j({k}/{2^j})$ for level $j=0,1,\dots$, and dyadic grid points ${k}/{2^j}$, $k=0,\dots,2^{j}-1$.

The Gaussian Haar-based multifractional process (GHBMP) was introduced by~\cite{ayache2025} by the following formula
\begin{equation}\label{eq1}
    X(t) \coloneqq \sum_{j=0}^{+\infty}  \sum_{k=0}^{2^{j}-1}
     \left(\int_{0}^{1} (t-s)_{+}^{H_{j,k}-{1}/{2}} h_{j,k}(s)ds \right)\varepsilon_{j,k}, \quad t \in [0,1],
\end{equation}
 where $\{\varepsilon_{j,k}\}$ denotes a sequence of independent $\mathcal{N}(0,1)$ Gaussian random variables,  $\{h_{j,k}(\cdot)\}$ are the Haar wavelets that can be represented using dyadic dilations and translations of the indicator functions $\mathbbm{1}_A (\cdot)$ of a set $A$ as 
 \begin{equation*}
 h_{j,k} (s) = 2^{j/2} \left({\mathbbm{1}}_{[k/2^j,(k+1/2)/2^j)}(s) -{\mathbbm{1}}_{[(k+1/2)/2^j,(k+1)/2^j)}(s)\right).
 \end{equation*}
 For numeric computations, the integrals in (\rm {\ref{eq1}}) can be simplified to
\begin{align}
        2^{-j H_{j,k}} h^{[H_{j,k}]} (2^jt-k) & \coloneqq \int_{0}^{1} (t-s)_{+}^{H_{j,k}-\frac{1}{2}} h_{j,k} (s) ds \nonumber\\
         & \hspace*{-1cm} \coloneqq  \dfrac{(2^jt-k)_+^{H_{j,k} + \frac{1}{2}} -2\left(2^jt-k-\frac{1}{2}\right)^{H_{j,k} + \frac{1}{2}}_+ + \left(2^jt-k-1\right)^{H_{j,k} + \frac{1}{2}}_+}{2^{j H_{j,k}}\left(H_{j,k} + \frac{1}{2}\right)}, \nonumber
\end{align}
where, $(x)_{+} \coloneqq \max(x,0)$, $x \in \mathbb{R}$. 

This process $X(t), t \in [0,1]$, is a Gaussian centered stochastic process, which is defined on the interval $[0,1]$.  It was proved in ~\cite{ayache2025} that, for the considered class of processes, by construction in (\ref{eq1}), their pointwise H\"older exponent $\alpha_X(t)$  equals the Hurst function 
$H(t),$ $t \in [0,1]$. A detailed discussion of the properties of GHBMP is given by~\cite{ayache2025}. A truncated version of the series in (\rm {\ref{eq1}}) was used to develop an R function to generate realizations of~GHBMP.

The theoretical covariance function of the GHBMP, \citep[see][]{ayache2025}, is given by
\begin{equation}\label{eq2}
{\rm Cov}(X(t),X(t'))=\sum_{j=0}^{+\infty}  \sum_{k=0}^{2^{j}-1} 2^{-2jH_{j,k}} h^{[H_{j,k}]} (2^{j}t-k) h^{[H_{j,k}]} (2^{j}t'-k).
\end{equation}
This function is useful in applications where not only simulated values of the processes are required, but also their dependence structure. In the package, a truncated version of formula (\rm {\ref{eq2}}) was used for computations, with a sufficient number of terms taken in the first sum.

\subsection{Estimation of Hurst and covariance functions} 

The Hurst function governs the local H\"{o}lder regularity of multifractional processes and time series. Accurate statistical estimation of the Hurst function is important for identifying the time-varying roughness and local behavior of processes. The approach proposed by \cite{ayache2022uniformly} was applied to obtain statistical estimates of the Hurst function. The estimator $\hat{H}(t)$ is uniformly and strongly consistent for $H(t)$. Algorithm~\ref{algo1} outlines the method and computational steps realized in the corresponding \pkg{Rmfrac} function. 

\begin{algorithm}[!htb]
\caption{Estimation of the Hurst function}
\label{algo1}
\begin{algorithmic}[0]
\State 1: \textbf{Input:} Realization of $X$, positive integers  $L \geq 2$, $Q \geq 2$ and $N$ 
\State 2: \textbf{for} each $l \in \{0,\dots,L\}$ compute
      $$a_{l} \coloneqq (-1)^{L-l} \dfrac{L!}{l!(L-l)!}$$  
\State 3: \textbf{for} each $k \in \{0, \dots N-L\}$ and $N$ compute
     $$d_{N,k} \coloneqq \sum_{l=0}^{L} a_{l} X\left(\frac{k+l}{N}\right)$$  
\State 4: Partition the interval $I_{N,n}$ into $N$ sub-intervals \\ 
\hspace{1.2em}$I_{N,n} \coloneqq \left[\frac{n-1}{N},\frac{n}{N}\right)$ for $n=1,\dots,N-1$ and $I_{N,N} \coloneqq \left[ \frac{N-1}{N},1\right]$
\State 5: \textbf{for} each $I_{N,n}$ and $k \in \nu_{N} (I) \coloneqq \{ k \in \{0, \cdots, N-L \}: \frac{k}{N} \in I \}$ compute the \\
\hspace{1.2em}generalized quadratic variations:
     $$V_{N} (I_{N,n}) \coloneqq |\nu_{N} (I_{N,n})|^{-1 } \sum_{k \in \nu_{N} (I_{N,n})} |d_{N,k}|^{2}$$  
\State 6: \textbf{for} each $I_{N,n}$ compute the estimate:
     $$\hat{H}_N^{Q} (I_{N,n}) \coloneqq \min \left( \max \left( \log_{Q^2} \left( \dfrac{V_{N}(I_{N,n})}{V_{QN}(I_{N,n})} \right) ,0 \right) ,1 \right)$$  
\State 7: \textbf{return} Hurst function estimates
\end{algorithmic}
\end{algorithm}

The relationship $\hat{D}(t)=2-\hat{H}(t)$ between the local fractal dimension $D(t)$ and the Hurst function $H(t)$ was used to estimate the local fractal dimension \citep{Frac_Dim_Tilmann}. 

The empirical covariance function of $X(t)$ provides an estimate of the true covariance structure based on multiple independent realizations of the process. For $M$ realizations $X^{(k)}(t),$ $k=1,...,M,$ of the process, the empirical covariance between values at the time points $t_i$ and $t_j$ is defined as 
\begin{equation*}
\hat{C}(t_i,t_j) \coloneqq \frac{1}{M} \sum_{k=1}^{M} (X^{(k)}(t_i)-\bar{X}(t_i))(X^{(k)}(t_j)-\bar{X}(t_j)),
\end{equation*}
where $\bar{X}(t)$ is the empirical mean of the realizations $X^{(k)}(\cdot)$ at the time point~$t$. Assuming the second moments of the process are finite and realizations are independent or weakly dependent, by the central limit theorem, the empirical covariance converges to the true one as $M \rightarrow \infty$. This ensures that the estimates are reliable when the number of realizations is sufficiently large.

\subsection{Cluster analysis of multifractional data}

Clustering processes involves grouping their realizations, so that the realizations within a cluster are more alike than realizations in other clusters. The main aim of the considered clustering is to identify realizations with similar roughness. We utilize the estimates of the local H\"{o}lder regularity of each realization to measure their similarities. The two popular clustering techniques, the hierarchical and k-means, are employed. The \code{hclust} and the \code{kmeans} functions available in the \pkg{stats} R package are used in developing the algorithms for hierarchical and k-means clustering, respectively. 

The Algorithm~\ref{algo2} shows how the hierarchical clustering (agglomerative) works in this context.

\begin{algorithm}[!htb]
\caption{Hierarchical clustering of realizations}
\label{algo2}
\begin{algorithmic}[0]
\State 1: \textbf{Input:} Realizations of the processes\\
                       \hspace{4.8em}Distance method (e.g. Euclidean, Manhattan, Minkowski)\\
                        \hspace{4.8em}Linkage criteria: ward.D, ward.D2, single, complete, average, mcquitty,\\
                         \hspace{4.8em}median and centroid\\
                        \hspace{4.8em}The number of clusters $k$ or the height $h$ at which the dendrogram is cut 
                         
\State 2: \textbf{for} each realization create
 \State \hspace{2.6em} a data frame of the smoothed Hurst function estimates (row-wise)  
\State 3: Assign each row of Hurst estimates as a singleton cluster
\State 4: Set  
$$cluster\_count = \text{number of rows in the data frame}$$
\State 5: \textbf{for} the chosen distance method compute
 \State \hspace{2.6em} the pairwise distance matrix of the singleton clusters (between rows of the\\
 \hspace{2.9em}data frame)  
\State 6: \textbf{for} the chosen linkage criteria \textbf{while}
\State \hspace{2.6em} $cluster\_count >1$ \textbf{do}
 \State \hspace{2.6em} merge the closest pair of clusters into a single cluster
 \State \hspace{2.6em} update the distance matrix using the linkage criteria
\State 7: Obtain the cluster assignments based on $k$ or $h$ 
\State 8: \textbf{return} Cluster dendrogram, list of realisations in each cluster, centres of \\ 
\hspace{0.78em} clusters, smoothed Hurst estimates, and other information as in Table~\ref{table4}
\end{algorithmic}   
\end{algorithm}

Similarly, the iterative Algorithm~\ref{algo3} realizes the k-means clustering. The centroid-based clustering technique is applied for a specified number of clusters $k$. 

\begin{algorithm}[!htb]
\caption{k-means clustering of realizations}
\label{algo3}
\begin{algorithmic}[0]
\State 1: \textbf{Input:} Realizations of the processes\\
                       \hspace{4.8em}Number of clusters $k$\\
                       \hspace{4.8em}Maximum number of iterations for each run $iter.max$\\
                       \hspace{4.8em}Number of runs $nstart$
\State 2: \textbf{for} each realization create
 \State \hspace{2.6em} a data frame of the smoothed Hurst function estimates (row-wise)  
\State 3: \textbf{for} $run = 1:nstart$ choose
\State \hspace{2.6em} $k$ initial random centroids ($k$ rows from the data frame) 
\State 4: \hspace{1.8em}\textbf{for} $iter = 1:iter.max$
  \State \hspace{4.8em}calculate the distance between centroids and each row (item) in the data 
 \State  \hspace{4.8em} frame
    \State \hspace{4.8em}assign each item to the nearest centroid
    \State \hspace{4.8em}update the cluster centroids as the mean of each item in the cluster    
    \State \hspace{4.8em}calculate the within cluster sum of squares \textit{WCSS}
\State  \hspace{3.0em}\textbf{end for}
 \State \hspace{3.0em}store the minimum \textit{WCSS} and cluster assignments for this $run$
\State \hspace{1.2em}\textbf{end for}
\State 5: Obtain the assignment of each smoothed Hurst function estimate (item) to its \\ 
\hspace{0.78em} nearest cluster from the run with the minimum WCSS
\State 6: \textbf{return} A list of realisations in each cluster, centres of clusters, smoothed Hurst \\ 
\hspace{0.78em} estimates and other information as in Table~\ref{table4}
\end{algorithmic}   
\end{algorithm}

\subsection{Geometric statistics}

Several statistical estimators were realized in the package. This section provides examples of three of them used in the numerical illustrations.

The sojourn measure is the total length of sub-intervals of the given time interval $[T_1,T_2]$, where a time series $X(t)$ stays above or under a certain constant level $A$. It can be estimated by the following statistics
\begin{equation}\label{eq3}
\hat{\mu}_{T}^+ (A) \coloneqq \delta \sum_{k=0}^{[T_2-T_1/\delta]} \mathbbm{1}_{\{X(T_1+k\delta) \geq A\}}  \quad \text{or} \quad \hat{\mu}_{T}^- (A) \coloneqq \delta \sum_{k=0}^{[T_2-T_1/\delta]} \mathbbm{1}_{\{X(T_1+k\delta) \leq A\}},
\end{equation}
respectively, where $\delta$ is a time step and $[\cdot]$ is the integer part function. 

The excursion area of a realization of a random process above or under a certain constant level $A$ can be estimated by
\begin{equation}\label{eq4}
 \hat{A}_{T}^{+} (A) \coloneqq \delta \sum_{k=0}^{[T_2-T_1/\delta]} (X(T_1 + k\delta)-A)_+   \quad \text{or} \quad \hat{A}_{T}^{-} (A) \coloneqq \delta \sum_{k=0}^{[T_2-T_1/\delta]} (A-X(T_1 + k\delta))_+ ,
\end{equation}
respectively. 

The Relative Strength Index (RSI) is a popular statistic (scales between $0$ and $100$) often used in financial applications to measure the extent of price changes and to evaluate overbought (RSI above $70$) or oversold (below $30$) values. It was introduced by~\cite{wilder1978new}. To compute RSI, the formula below is applied
\begin{equation*}
RSI = 100 - \dfrac{100}{1+RS},
\end{equation*}
where $RS$ denotes the relative strength, which is given by the proportion of average gains and average losses over a specified period $n$ (in many financial applications $n=14$ is used). Initial values of the average gain and average loss are obtained by the arithmetic mean over the selected period. Then, Wilder’s smoothing method is used according to the following formulas for the subsequent periods to update their values recursively,
\begin{equation*}
Avg\_gain_t = \frac{(Avg\_gain_{t-1} \times (n-1)) + Gain_t}{n},
\end{equation*}
\begin{equation*}
Avg\_loss_t = \frac{(Avg\_loss_{t-1} \times (n-1)) + Loss_t}{n}.
\end{equation*}

\section{\pkg{Rmfrac} package} \label{sec3}

The \pkg{Rmfrac} R package \citep{rmfrac} can be installed from CRAN. Its most recent development version is also available on GitHub (\url{https://github.com/Nemini-S/Rmfrac}). 

\begin{figure}[!htb]
\begin{center}

\begin{tikzpicture}[
    box/.style={
        rectangle, draw, minimum width=3.3cm, minimum height=1.7cm, align=left, text width=3.1cm
    },
    bigbox/.style={
        rectangle, draw, minimum width=4.5cm, minimum height=5.4cm, align=left, text width=4.1cm
    },
    ->, >=Stealth
]

\node[bigbox] (box1) {
    \textbf{Analysis and \\estimation} \\
    \rule{\linewidth}{0.4pt} \\
    \textbf{Covariance functions} \\
    \vspace{0.1cm}
    \begin{tabular}{l}
        \code{est\_cov} \\
        \code{cov\_GHBMP}
    \end{tabular}
    \rule{\linewidth}{0.4pt} \\
    \textbf{Clustering} \\
    \vspace{0.1cm}
    \begin{tabular}{l}
        \code{hclust\_hurst}\\
        \code{kmeans\_hurst}\\
        \code{print.hc\_hurst}\\
        \code{print.k\_hurst}
    \end{tabular}
    \rule{\linewidth}{0.4pt} \\
    \textbf{Geometric statistics} \\
    \vspace{0.15cm}
    \begin{tabular}{l}
        \code{sojourn}\\
        \code{exc\_Area}\\
        \code{RS\_Index}\\
        \code{long\_streak}\\
        \code{mean\_streak}\\
        \code{cross\_T}\\
        \code{cross\_rate}\\
        \code{cross\_mean}\\
        \code{X\_max}\\
        \code{X\_min}
    \end{tabular}
};

\node[box, right=4.6cm of box1.north, anchor=north] (box2) {
    \textbf{Shiny app} \\
    \vspace{0.1cm}
    \begin{tabular}{l}
     \code{shinyapp\_sim}
  \end{tabular}
};

\node[box, right=4.4cm of box2.north, anchor=north] (box3) {
    \textbf{\mbox{Main estimators}} \\
    \vspace{0.2cm}
    \begin{tabular}{l}
        \code{Hurst}\\
        \code{LFD}\\
        \code{H\_LFD}
    \end{tabular}
};

\node[box, below=1.5cm of box2] (box4) {
    \textbf{Simulation of processes} \\
    \vspace{0.2cm}
    \begin{tabular}{l}
     \code{Bm}\\
     \code{Bbridge}\\
      \code{FBm}\\
      \code{FBbridge}\\
      \code{FGn}\\
     \code{GHBMP}
    \end{tabular}
};

\node[box, below=3.3cm of box3] (box5) {
    \textbf{Visualizations} \\
    \vspace{0.2cm}
    \begin{tabular}{l}
     \code{plot.mp}\\
     \code{plot.H\_LFD}\\
     \code{plot\_tsest}\\
     \code{plot.hc\_hurst}\\
     \code{plot.k\_hurst}\\
  \end{tabular}    
};

\draw[->] (box4.west) -- (box1.east|-box4.west);
\draw[->] (box3.west) -- (box2.east|-box3.west);
\draw[->] (box4) -- (box2);
\draw[->] (box3) -- (box5);
\draw[->] (box4) -- (box5);
\draw[->] (box4) -- (box3);
\draw[->] ([yshift=-11mm]box1.east|-box5.west) -- ([yshift=-11mm]box5.west);
\draw[->] ([yshift=3cm]box3) -- (box1);
\draw[->] (box1.east|-box2.west) -- (box2.west);
\draw[->] (box5) -- (box2);
\end{tikzpicture}
\end{center}
\caption{Structure of the \pkg{Rmfrac} package}
\label{fig1}
\end{figure}
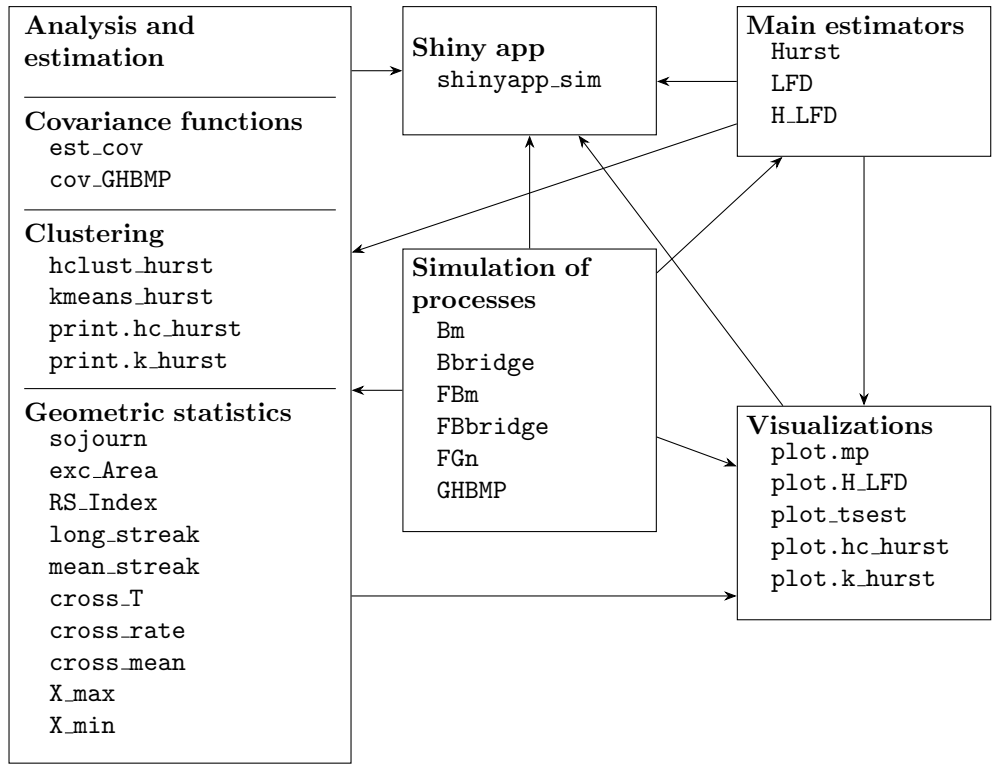

The functions in \pkg{Rmfrac}  package can be categorized into three main groups: simulation of processes, estimation of the Hurst function and local fractal dimension, and analyses of processes and time series. There is also a group of plotting functions for each class of outputs from the main groups. Figure~\ref{fig1} visualizes the structure of the package, the main functions in each category and their relationships.  

This section uses the following time sequences to demonstrate the functionality of the package.
\begin{lstlisting}
t1 <- seq(0, 1, by = (1/2)^14)
t2 <- seq(0, 1, by = 0.01)
t3 <- seq(0, 1, by = (1/2)^11)
\end{lstlisting}

\subsection{Simulation of processes and estimation of Hurst functions and local fractal dimension} \label{sec3.1}

\begin{figure}
    \centering
    \includegraphics[width=1\linewidth]{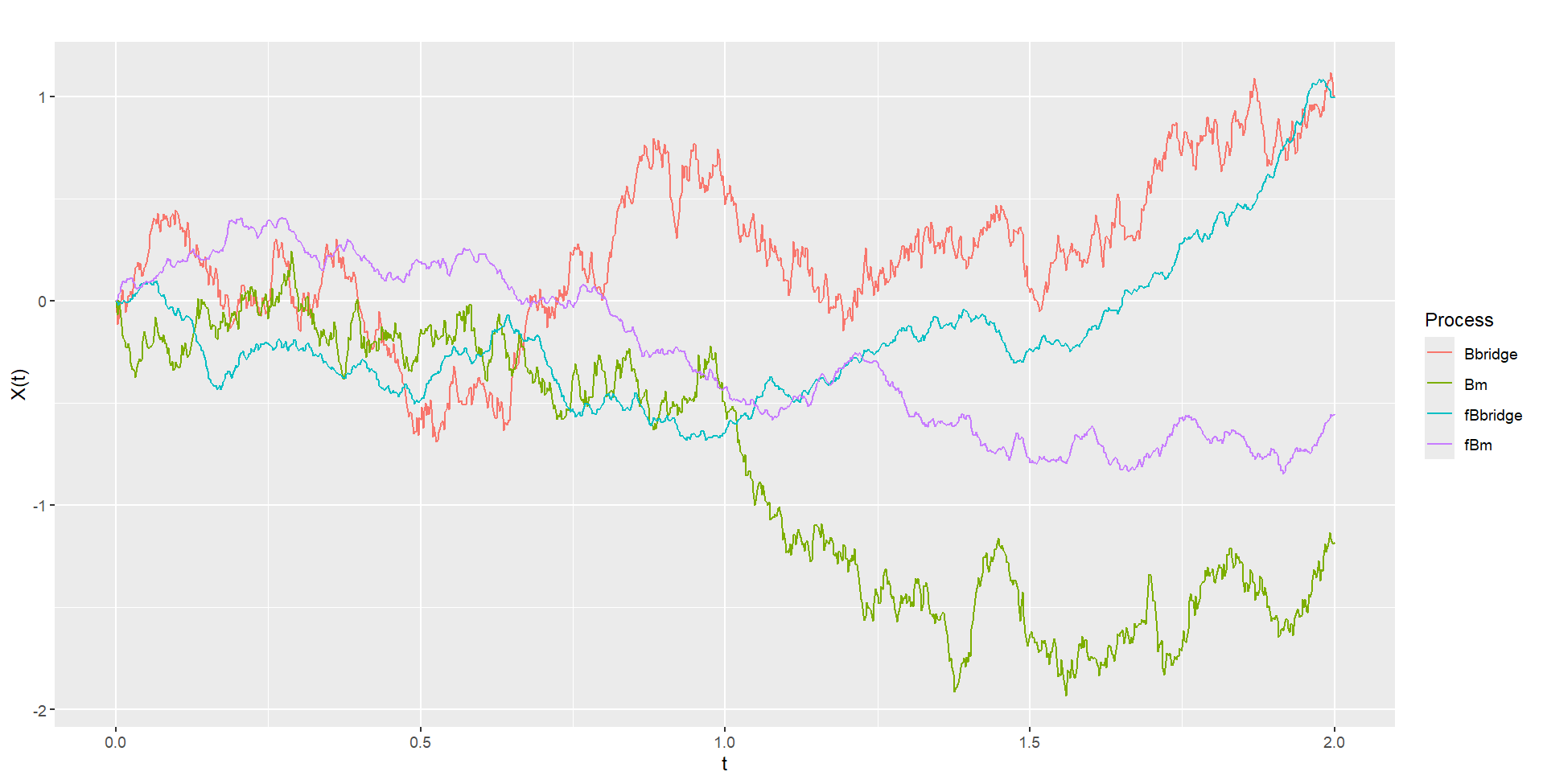}
    \caption{Realizations of \code{Bm}, \code{Bbridge}, \code{FBm} and \code{FBbridge}}
    \label{fig2}
\end{figure}

The simulation group includes six functions for simulating realizations of Bm, Brownian bridge, fBm, fractional Brownian bridge, fractional Gaussian noise and Gaussian Haar-based multifractional processes. The first five functions are included for completeness and to ensure consistent usage of these models. Figure~\ref{fig2} shows the realizations of the first four processes from functions available in the \pkg{Rmfrac} package. These are simulated over the time interval $[0,2]$ using $1000$ time steps.

The \code{GHBMP} function is used to simulate a Gaussian Haar-based multifractional process at a given time point or any time sequence in the interval $[0,1]$. The arguments for \code{GHBMP} are provided in Table~\ref{table1}. This function returns an \code{S3} object of class \code{"mp"} which includes the \code{plot} generic method. The object is a data frame with each row including a time point and the corresponding simulated value.  
\begin{table}[htb!]
\centering
\caption{Arguments of \code{GHBMP} function}
\label{table1}
\begin{tabular}{@{}l l@{}} 
\toprule
 Argument & Description \\ 
 \midrule
 \code{t} &  A time point or time sequence from the interval $[0,1]$. \\

 \code{H} &  Hurst function, defined as a function of \code{t}. \\ 
 
 \code{J} & The level at which the first summation is truncated in (\ref{eq1}). Default is $15$.\\ 

 \code{num.cores} & The number of cores to initiate a cluster for parallel computing. By default a cluster \\
& that uses all available cores minus one is created.\\
\botrule
\end{tabular}
\end{table}

To illustrate the usage of the \code{GHBMP} function, we consider several examples. When simulating a multifractional process with a constant Hurst function note that the constant must be defined as a function as well. The following examples illustrate the simulation of a multifractional process with a constant Hurst function of $0.3$ and two varying Hurst functions, namely, the linear function $0.8 - 0.55t$ and the oscillating function $0.4-0.25\sin (6\pi t)$.
\begin{lstlisting}
#Constant Hurst function
H1 <- function(t) {return(0.3 + 0 * t)}
Process_1 <- GHBMP(t1, H1, J = 14)
\end{lstlisting}
\begin{lstlisting}
#Linear Hurst function
H2 <- function(t) {return(0.8 - 0.55 * t)}
Process_2 <- GHBMP(t1, H2, J = 14)
\end{lstlisting}
\begin{lstlisting}
#Oscillating Hurst function
H3 <- function(t) {return(0.4 - 0.25 * sin(6 * pi * t))}
Process_3 <- GHBMP(t1, H3, J = 14)
\end{lstlisting}
To illustrate a simulation of a multifractional process with a piecewise Hurst function, the following \code{ifelse} syntax was used to define the Hurst function. 

\begin{lstlisting}
J <- 14
H4 <- function(t) {
   ifelse(t >= 0 & t <= (0.5 - 1/(2 * J)), 0.2,
   ifelse(t > (0.5 - 1 / (2 * J)) & t <= (0.5 + 1 / (2 * J)), 
        (0.6 * J * t) + (0.5 - 0.3 * J), 
   ifelse(t > (0.5 + 1 / (2 * J)) & t <= 1, 0.8, 0)))}
Process_4 <- GHBMP(t1, H4, J = 14)
\end{lstlisting}
Note that, when $J \to \infty,$ the Hurst function \code{H4} of \code{Process\_4} converges to the following discontinuous function
\begin{equation}\label{eq102}
    \lim_{J \to \infty} H_{J}(t) = H(t) = 
     \begin{cases}
        0.2, & \text{if } t \in \left[0,{1}/{2}\right]; \\
        0.8, & \text{if } t \in \left({1}/{2},1\right].
    \end{cases}
\end{equation}

The functions \code{Hurst} and \code{LFD} in \pkg{Rmfrac} provide the statistical estimates of the Hurst function and local fractal dimension, respectively. See Table~\ref{table2} for the usage of arguments of these two functions. Note that these functions can be applied to realizations of any random process or time series. Since these are estimators of local characteristics, to obtain reliable results, it is recommended to use a sufficiently large number of data points (at least 500). The return is a data frame containing the estimates obtained for $N$ sub-intervals of the time interval, with two columns denoting the time and the corresponding estimate.
\begin{table}[htb!]
\centering
\caption{Arguments of \code{Hurst} and \code{LFD} function}
\label{table2}
\begin{tabular}{@{}l l@{}} 
\toprule
 Argument & Description \\ 
\midrule
 \code{X} & A data frame with values of a time series, where the first column is the time sequence\\  
 &  and the second one contains values of the time series.\\

 \code{N} &  Number of sub-intervals the estimation to be performed on. By default \code{N} is $100$.\\ 
 
 \code{Q}\footnotemark[1] & Fixed integer greater than or equal to 2. By default \code{Q} is $2$.\\ 

 \code{L}\footnotemark[1] & Fixed integer greater than or equal to 2. By default \code{L} is $2$.\\
\botrule
\end{tabular}
\footnotetext[1]{The integer parameters \code{Q} and \code{L} define the generalized quadratic variations and the corresponding estimator, see Algorithm~\ref{algo1} and further mathematical details in  \cite{ayache2022uniformly}.}
\end{table}

The following examples illustrate the usage of \code{Hurst} and \code{LFD} to obtain the estimates of the Hurst function and local fractal dimension of \code{Process\_1}.

\begin{lstlisting}
head(Process_1)    
        t           X
1 0.000000e+00  0.00000000
2 6.103516e-05 -0.07667098
3 1.220703e-04 -0.14072544
4 1.831055e-04 -0.12711011
5 2.441406e-04 -0.15125273
6 3.051758e-04 -0.19429584
\end{lstlisting}
\begin{lstlisting}
Est_H <- Hurst(Process_1, N = 100)
head(Est_H)     
  Time Hurst_estimate
1 0.00      0.2912691
2 0.01      0.3613973
3 0.02      0.0000000
4 0.03      0.1448155
5 0.04      0.4430869
6 0.05      0.2536086
\end{lstlisting}
\begin{lstlisting}
Est_LFD <- LFD(Process_1, N = 100)
head(Est_LFD)     
  Time LFD_estimate
1 0.00     1.708731
2 0.01     1.638603
3 0.02     2.000000
4 0.03     1.855184
5 0.04     1.556913
6 0.05     1.746391
\end{lstlisting}

\begin{figure}[t!]
\centering
\begin{subfigure}[c]{0.375\textwidth}
\centering
    \includegraphics[trim=0 0 0 0, clip, width=1.1\textwidth, height= 1.3\textwidth]{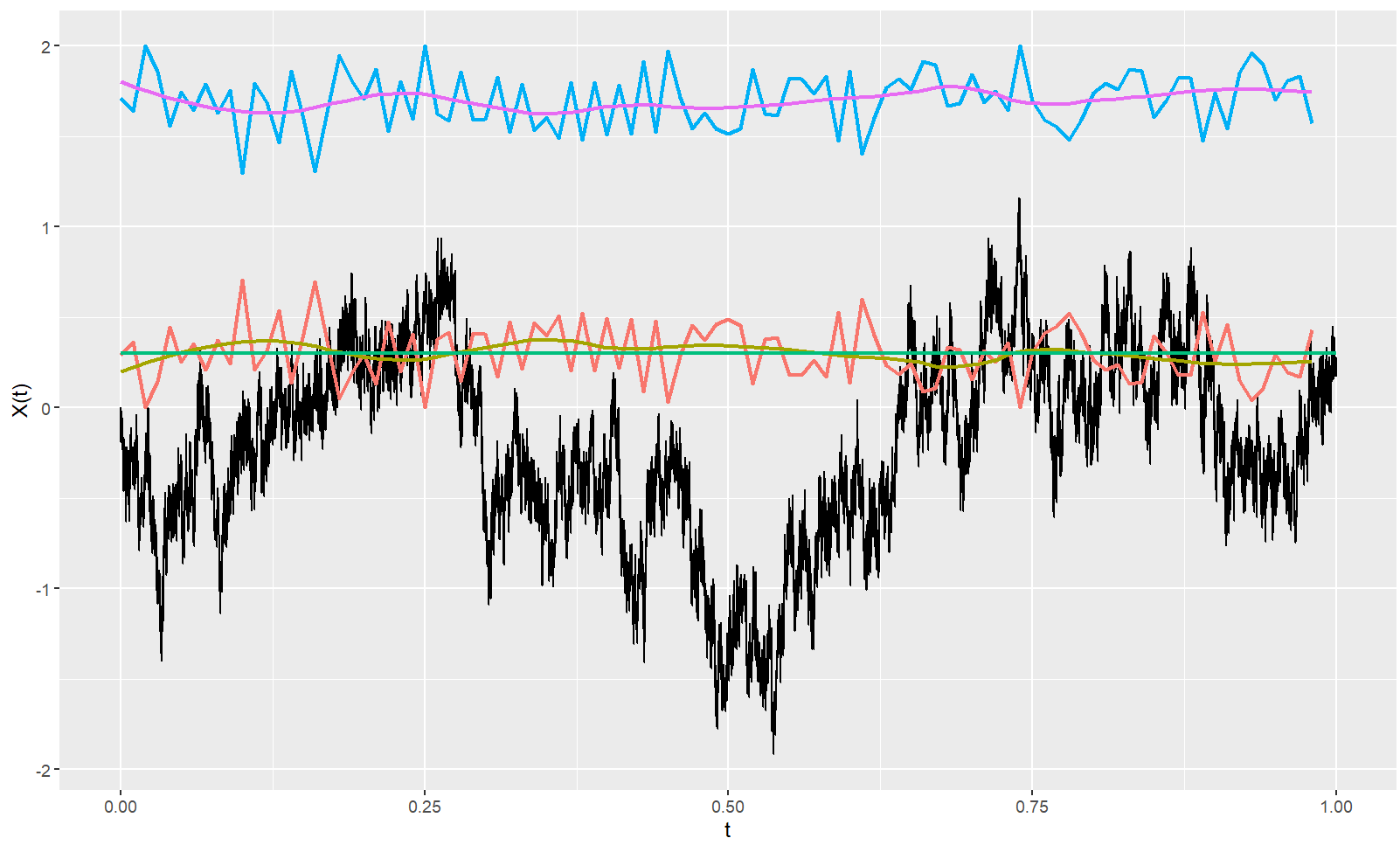}
    \caption{\mbox{\code{Process\_1}, $H\equiv 0.3$}}
\end{subfigure}\hspace{4mm}
\begin{subfigure}[c]{0.516\textwidth}
\centering
    \includegraphics[trim=0 0 0 1mm, clip,width=1.05\textwidth, height= 0.93\textwidth]{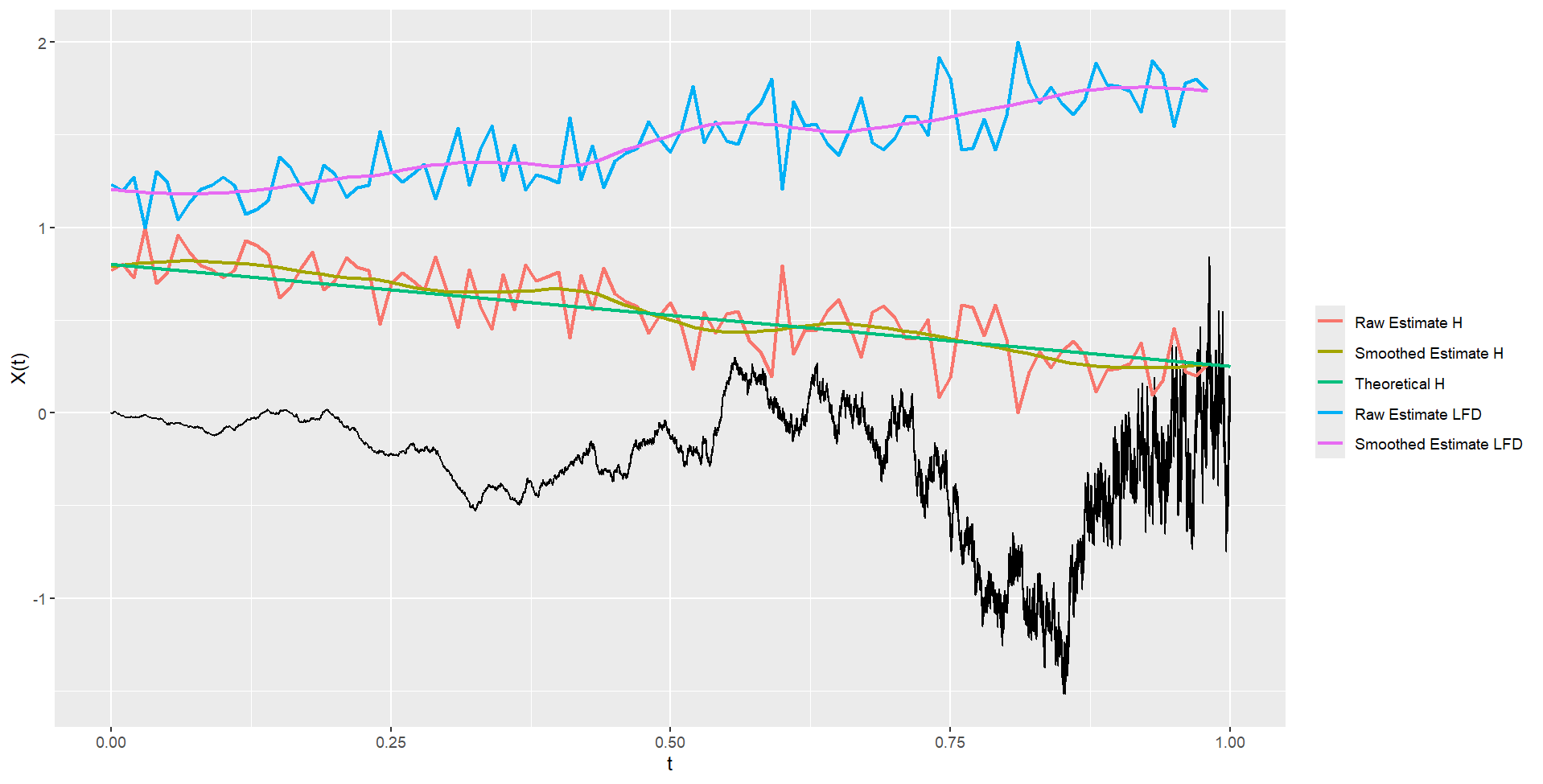}
    \caption{\mbox{\code{Process\_2}, $H(t)=0.8-0.55t$}}
\end{subfigure}

\begin{subfigure}[c]{0.39\textwidth}
\centering
    \includegraphics[trim=0 0 0 0, clip, width=1.1\textwidth, height= 1.34\textwidth]{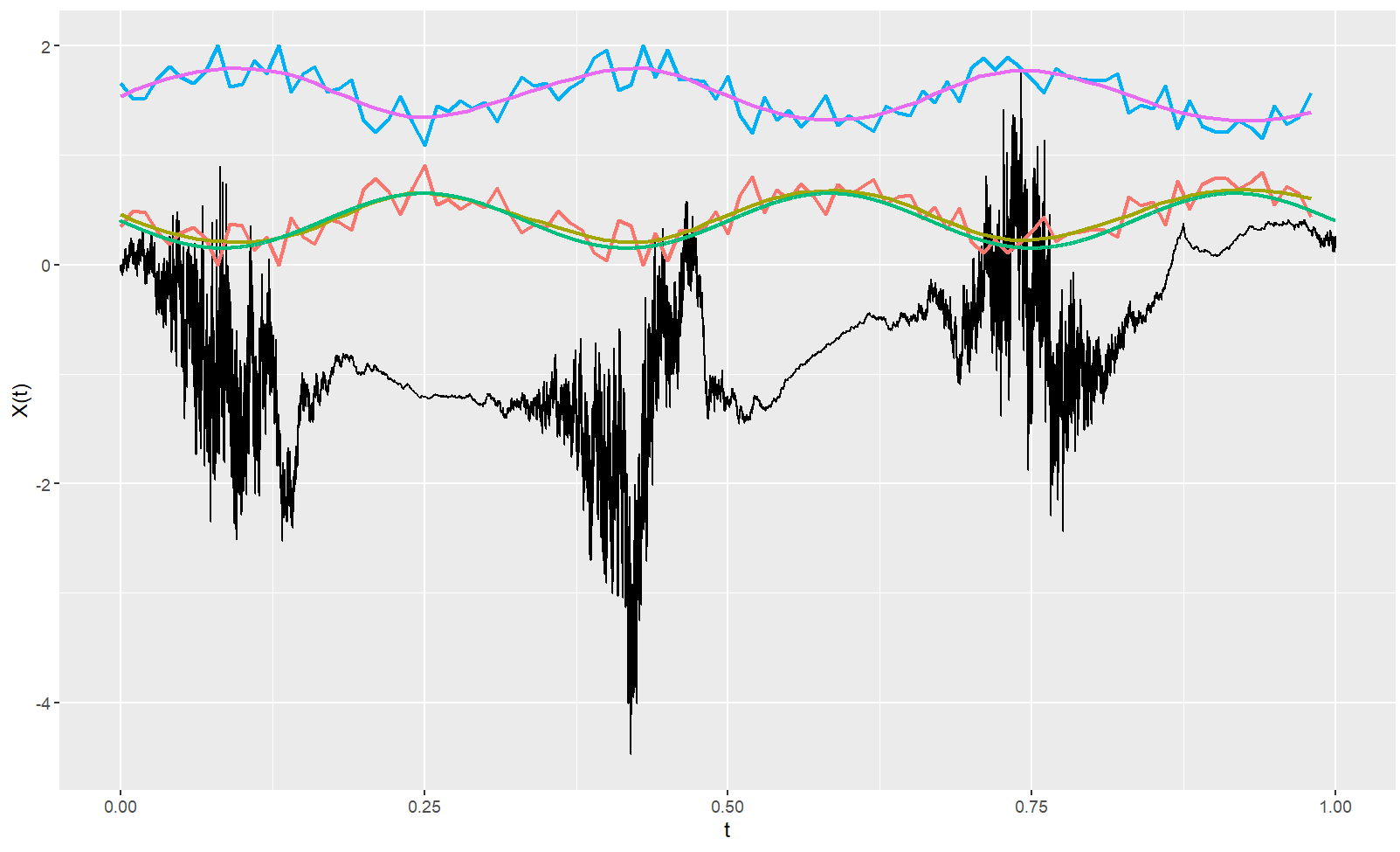}
    \caption{\mbox{\code{Process\_3}, $H(t)=0.4-0.25\sin(6\pi t)$}}
\end{subfigure}\hspace{4mm}
\begin{subfigure}[c]{0.527\textwidth}
\centering
    \includegraphics[trim=0 0 0 1mm, clip,width=1.05\textwidth, height= 0.98\textwidth]{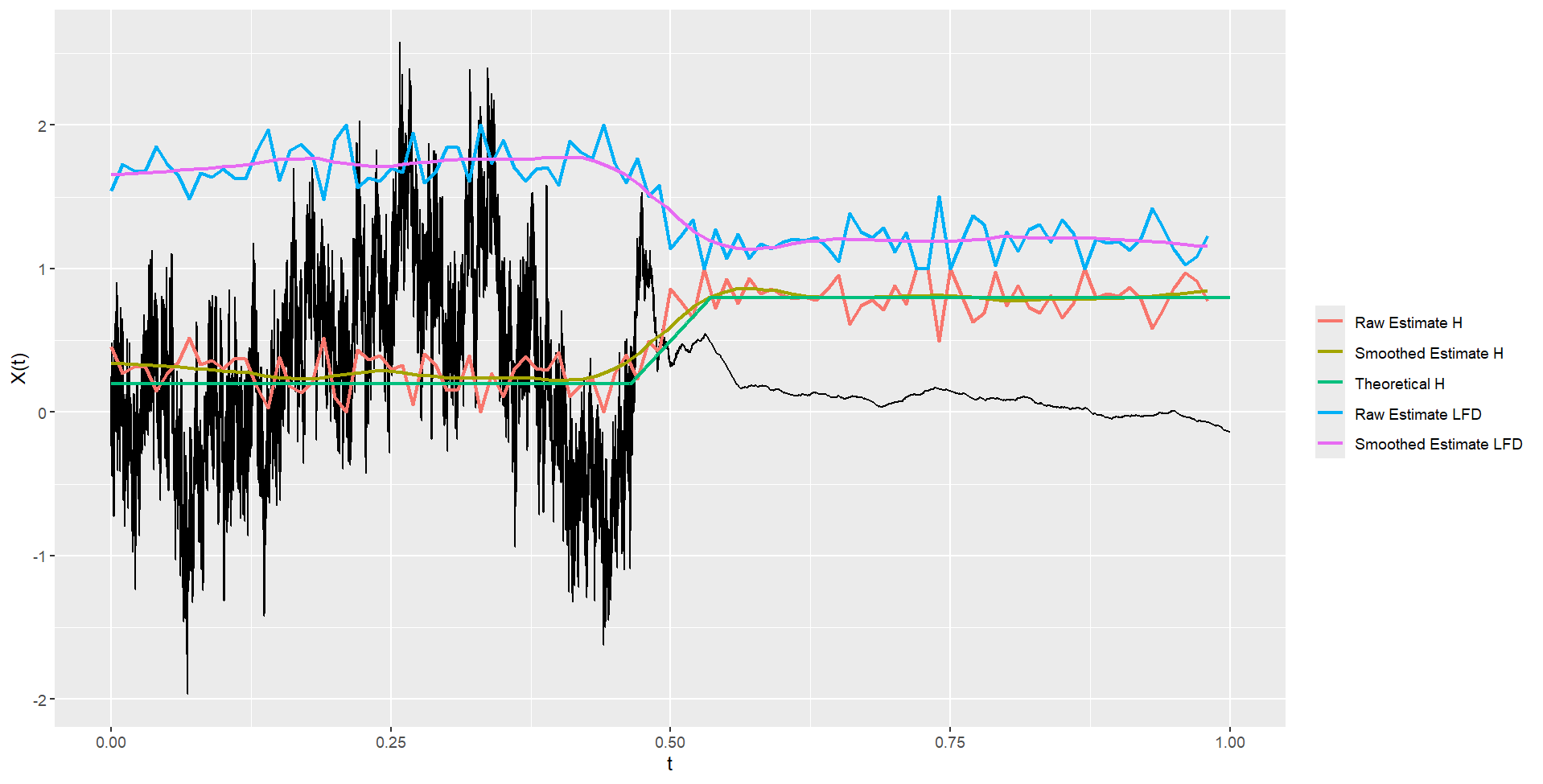}
    \caption{\code{Process\_4}, $H(t)$ is a piecewise function}
\end{subfigure}
\caption{Realizations of \code{GHBMP} and corresponding Hurst functions}
\label{fig3}
\end{figure}

The \code{S3} plot method for the \code{GHBMP} function provides the options to plot the simulated Gaussian Haar-based multifractional process together with raw and smoothed estimates of the Hurst function and local fractal dimension. To obtain smoothed estimates, the LOESS smoothing method \citep{cleveland1979robust} was used. Moreover, this function has the option to plot the theoretical Hurst function used to simulate the GHBMP. The function works so that the return from \code{GHBMP} is utilized for estimation by \code{Hurst} and \code{LFD} (using the arguments \code{N}, \code{Q} and \code{L}). Then, the estimates are plotted with the realization of the process. Figure~\ref{fig3} includes the plots of \code{Process\_1}, \code{Process\_2}, \code{Process\_3} and \code{Process\_4} with their corresponding Hurst functions and fractal dimensions obtained using this \code{plot} method. For example, for \code{Process\_1} with the theoretical Hurst function \code{H1}, it was called as follows.

\begin{lstlisting}
plot(Process_1, H = H1, H_Est = TRUE, H_Smooth_Est = TRUE, LFD_Est = TRUE, LFD_Smooth_Est = TRUE)   
\end{lstlisting}  

The package also provides two functions to visualize a user-provided random process or time series with an overlay of raw and smoothed estimates of the Hurst function and local fractal dimension. The first function is
\begin{lstlisting}
plot_tsest(X, H_Est = TRUE, H_Smooth_Est = TRUE, LFD_Est = TRUE, LFD_Smooth_Est = TRUE, N = 100, Q = 2, L = 2)   
\end{lstlisting}   
where \code{X} is the user-provided process or time series. The other is an \code{S3} plot method for objects of class \code{H\_LFD} returned from the function \code{H\_LFD}. It has the same list of arguments, but \code{X} is replaced with the return from \code{H\_LFD}. Note that the \code{H\_LFD} function returns an object list of raw and smoothed Hurst and fractal dimension estimates, allowing the users to retrieve the underlying numerical results for further analysis.  

\begin{figure}[htb!]
  \centering
  \begin{subfigure}[b]{0.5\textwidth}
    \includegraphics[width=\textwidth,height=6.5cm]{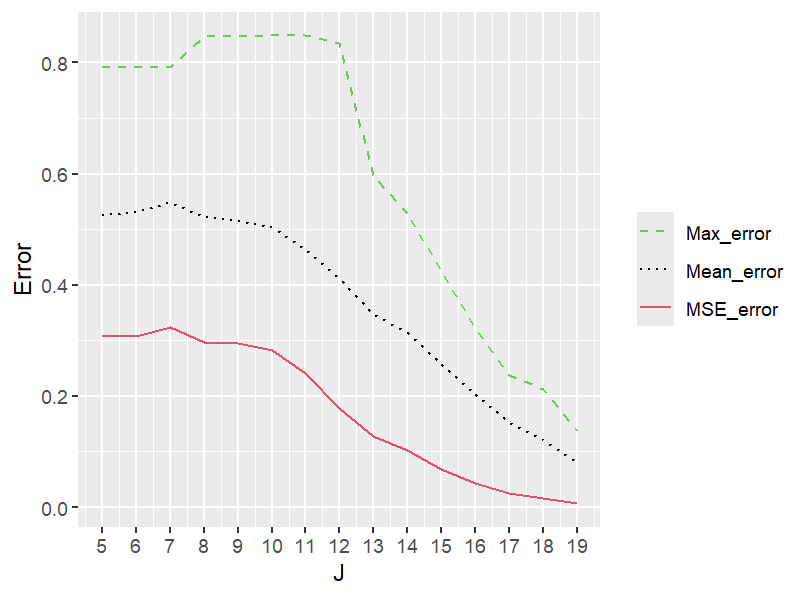}
    \label{fig_error_1}
  \end{subfigure}
   \hspace{0.01\textwidth}
  \begin{subfigure}[b]{0.4\textwidth}
    \includegraphics[width=\textwidth,height=6.5cm]{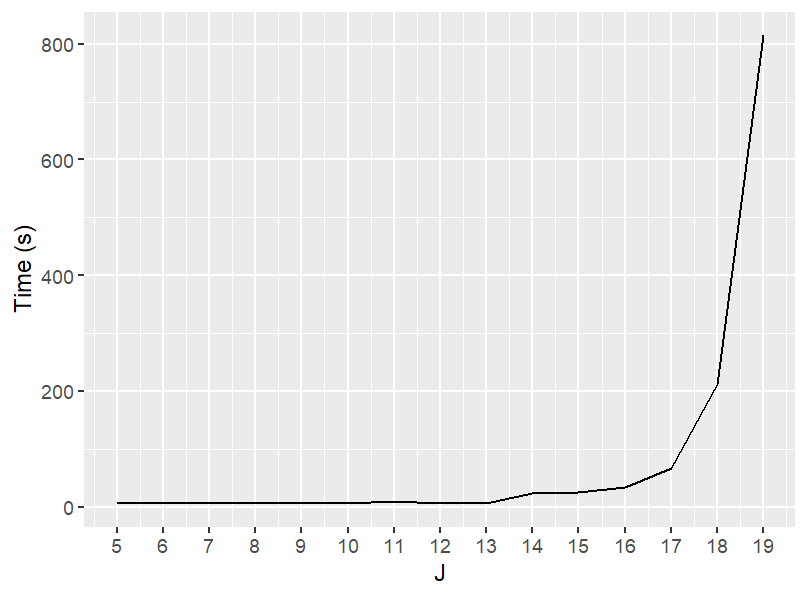}
    \label{fig_error_2}
  \end{subfigure}
  \caption{Change of average errors (left) and elapsed time (right) of \code{GHBMP} }
  \label{fig_errors}
\end{figure}

To assess the trade-off between computational efficiency and accuracy of the main function,  \code{GHBMP}, the errors and running time were compared for various values of \code{J}. For each \code{J},  the same number of simulations was performed. The time sequence from $0$ to $1$ with step size $(1/2)^n$ and Hurst function~$H(t)=0.4-0.25\sin(6\pi t)$ were used. For each realization, the maximum error, the mean error, and the mean squared error were computed from the difference between the theoretical Hurst function and its estimates obtained using \code{Hurst}. Table~\ref{table_errors} presents the average values of these errors and the average time elapsed (in seconds) over the simulations. For each \code{J}, a fixed value $n=\max(J-4,3)$  was chosen to provide enough time points for the computations. Figure~\ref{fig_errors} visualizes the average errors and elapsed times with \code{J}. The maximum error, mean error, and the mean squared error decrease with \code{J}. However, for small \code{J}, the simulation results are inaccurate, and the errors are almost identical. The mean error and the mean squared error tend to decrease more rapidly from around $\code{J}=10$ and the maximum error from $\code{J}=12$. The computational time starts to sharply increase at approximately $\code{J}=17$. As expected, the accuracy of the simulations increases with the running time. To balance the running time and accuracy, we selected the default value $\code{J}=15$ for the \code{GHBMP} function.

\begin{table}[htb!]
\centering
\caption{Average errors and computation time of \code{GHBMP}}
\label{table_errors}
\begin{tabular}{@{}c c c c c c@{}}
\toprule
$J$ &  Maximum error &  Mean error  & Mean squared error & Mean elapsed time\\
\midrule
5  &   0.7926283 & 0.52572095 & 0.307273518 & 6.124609 \\
6  &   0.7926283 & 0.53141051 & 0.308243796 & 6.101760 \\
7  &   0.7926283 & 0.54939139 & 0.323866724 & 6.144913 \\
8  &   0.8480287 & 0.52224806 & 0.297747087 & 6.229948 \\ 
9  &   0.8480287 & 0.51537352 & 0.294851220 & 6.232391 \\
10  &   0.8495067 & 0.50385490 & 0.282098100 & 6.273000 \\
11  &   0.8495067 & 0.46510440 & 0.240562690 & 9.615800 \\
12  &   0.8342982 & 0.41185070 & 0.177668250 & 7.165000 \\
13  &   0.5956134 & 0.34817160 & 0.127573090 & 6.333400 \\
14  &   0.5297822 & 0.31539250 & 0.103405080 & 23.427900 \\
15  &   0.4254869 & 0.25663850 & 0.068864770 & 25.852500 \\
16  &   0.3215763 & 0.20277520 & 0.042883160 & 33.533700 \\
17  &   0.2375536 & 0.15270380 & 0.024773230 & 67.751800 \\
18  &   0.2116933 & 0.12071450 & 0.015835060 & 212.763700 \\
19  &   0.1371554 & 0.08059045 & 0.006858024 & 815.037200 \\
\botrule
\end{tabular}
\end{table}

\subsection{Analyses of multifractional processes} \label{sec3.2}

This subsection discusses some functions from the \textit{analysis group} in the \pkg{Rmfrac} package that identify the covariance structure of a multifractional process, perform clustering of realizations and various associated estimates. 

To compute the theoretical covariance function of the Gaussian Haar-based multifractional process, use the \code{cov\_GHBMP} function. Its input arguments are similar to the arguments of the \code{GHBMP} function presented in Table~\ref{table1} with the default value of \code{J} set to $8$. In addition to these arguments, two further arguments are available. The first is an optional smoothing parameter \code{theta} to reduce noise and local irregularities of the covariance matrix. This was implemented using the \code{image.smooth} function from the \pkg{fields} package in R \citep{fields_R}. The second is a logical argument to plot the 3D surface plot of the theoretical covariance function. 

Notice that the theoretical covariance function of the process given by (\ref{eq2}) is positive definite.
When it is computed in the package by using the function \code{cov\_GHBMP}, a truncated version of the series in (\ref{eq2}) is used, but this truncation corresponds to the truncated series (\ref{eq1}) of the process. Therefore, the function \code{cov\_GHBMP} preserves the positive definiteness property. Also, due to the wavelet series representation with a large number of terms in (\ref{eq2}), the difference between the result of \code{cov\_GHBMP} and the true covariance $\mathrm{Cov}(X(t),X(t'))$ is very small. 

\begin{figure}[htb!]
  \centering
  \begin{subfigure}[b]{0.48\textwidth}
    \includegraphics[width=\textwidth,height=6cm]{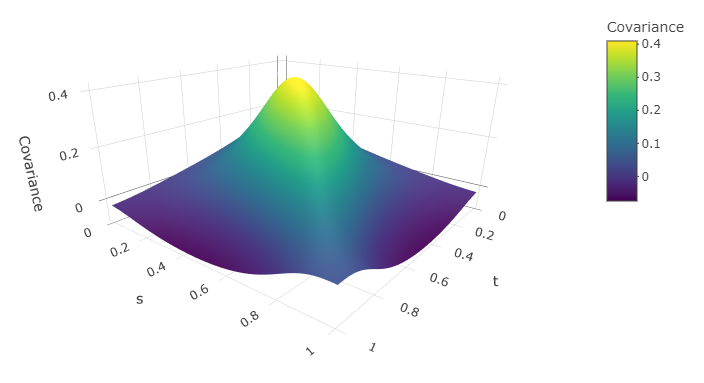}
    \caption{Theoretical covariance function}
    \label{fig41}
  \end{subfigure}
   \hspace{0.01\textwidth}
  \begin{subfigure}[b]{0.48\textwidth}
    \includegraphics[width=\textwidth,height=6cm]{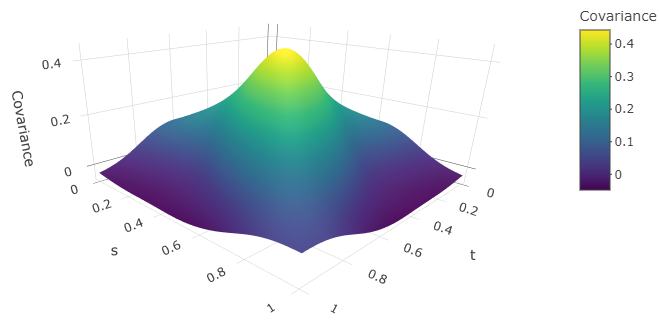}
    \caption{Smoothed estimated covariance function}
    \label{fig42}
  \end{subfigure}
  \caption{Theoretical (left) and estimated (right) covariance functions for GHBMP with Hurst function equal to $0.3$}
  \label{fig4}
\end{figure}

The \code{cov\_GHBMP} function returns a covariance matrix for each pair of time points in the provided time sequence. For example, to obtain the covariance matrix and the plot of the covariance function of a Gaussian Haar-based multifractional process with the Hurst function equal to $0.3$ use the \code{cov\_GHBMP} as:
\begin{lstlisting}
cov_GHBMP(t2, H1, theta = 0.1, plot = TRUE)   
\end{lstlisting}

Empirical covariance function of a process or time series can be computed using the \code{est\_cov} function. This function provides estimates for each pair of time points by using \code{M} realizations of the process or time series. The input should be a data frame with a time sequence in the first column and realizations in the remaining columns. The function also includes a smoothing parameter \code{theta} and a logical option to plot a 3D surface plot of the empirical covariance function. The return is a covariance matrix that gives the empirical covariance estimates between each pair of time points. Note that the time points that correspond to rows and columns in the matrix are arranged in ascending order. The following code chunk shows an example of computing the empirical covariance estimates of a GHBMP with the Hurst function equal to $0.3$ by using $20$ realizations. 
\begin{lstlisting}
X.t <- replicate(20, GHBMP(t2, H1), simplify = FALSE)
X <- do.call(rbind, lapply(X.t, function(df) df[, 2]))
Data <- data.frame(t2, t(X))
est_cov(Data, theta = 0.15, plot = TRUE)   
\end{lstlisting}
Figure~\ref{fig4} demonstrates close agreement between the theoretical covariance function and its empirical counterpart estimated from the simulated data.  

The \pkg{Rmfrac} package provides functionality for grouping realizations of random processes or time series utilizing their Hurst functions. For reliable results, use realizations with at least 500 data points. We consider examples of the \code{hclust\_hurst} function, which performs hierarchical clustering of realizations into a nested tree based on their similarity. Secondly, \code{kmeans\_hurst} will illustrate the clustering of realizations into a prespecified number of clusters.

Refer to Table~\ref{table3} for the usage of \code{hclust\_hurst} function. Provide either \code{k} or \code{h} to specify the number of clusters. If both are provided \code{k} will be used. The available \code{dist.methods} are \code{Bhjattacharyya}, \code{Bray}, \code{Canberra}, \code{Chord}, \code{divergence}, \code{Euclidean}, \code{fJaccard}, \code{Geodesic}, \code{Hellinger}, \code{Kullback}, \code{Levenshtein}, \code{Mahalanobis}, \code{Manhattan}, \code{Minkowski}, \code{Podani}, \code{Soergel}, \code{supremum}, \code{Wave} and \code{Whittaker}. A custom distance measure can be defined within the \code{proxy} package, and then used for \code{dist.method}. The available linkage criteria are, \code{"ward.D"}, \code{"ward.D2"}, \code{"single"}, \code{"complete"}, \code{"average"}, \code{"mcquitty"}, \code{"median"} and \code{"centroid"}. For further details of these criteria, refer to the \code{hclust} function documentation in the \pkg{stats} base package \citep{Rmanual2025}.
 
\begin{table}[htb!]
\centering
\caption{Arguments of \code{hclust\_hurst} function}
\label{table3}
 \begin{tabular}{@{}l l@{}} 
\toprule
 Argument & Description \\ 
\midrule
 \code{X.t} &  List of realizations. Each realization is given as a data frame where the first column \\
 & is a time sequence and the second is the values of the process.\\  

 \code{k} &  The desired number of clusters.\\

 \code{h} &  The height at which the dendrogram should be cut off.\\

 \code{dist.method} &  Any registered distance method in \code{dist} function available in the \pkg{proxy} R package.\\
 & Default is \code{"euclidean"}.\\
  
 \code{method} &  The linkage criteria for hierarchical clustering. Default is \code{"complete"}.\\

 \code{dendrogram} &  Logical option to plot the dendrogram.\\

 \code{N, Q, L} &  Parameters for estimation of the Hurst function of each realization. Refer Table~\ref{table2}\\
 & for description.\\
 
\botrule
\end{tabular}
\end{table}

The function \code{hclust\_hurst} returns a \code{S3} object of class \code{"hc\_hurst"}, which includes the \code{print} and \code{plot} generic methods. This object is a list comprising the components outlined in Table~\ref{table4}. The \code{S3} plot method creates a plot of the smoothed Hurst functions of each realization in each cluster with the corresponding cluster centers. The usage is \code{plot(x, type = "estimates")}. If only the estimates of the Hurst function need to be plotted, provide the argument \code{type = "estimates"}, for only the cluster centers of the estimates use \code{type = "centers"} and for both provide \code{type = "ec"}.

\begin{table}[htb!]
\centering
\caption{Return values of \code{hclust\_hurst} and \code{"k\_hurst"} functions}
\label{table4}
\begin{tabular}{@{}l l@{}} 
 \toprule
 Component & Description \\ 
 \midrule
 \code{cluster\_info} & The cluster number of each smoothed estimated Hurst function (item) and\\
 & distances to their cluster centers.\\  

 \code{cluster} & Cluster number of each item.\\  

 \code{cluster\_sizes} & Number of items in each cluster.\\  

 \code{centers} & Cluster centers.\\  
  
 \code{smoothed\_Hurst\_estimates} & Smoothed Hurst estimates of each realization.\\  

 \code{raw\_Hurst\_estimates} & Raw Hurst estimates for each realization.\\  
 \code{call} & Information about the input parameters used.\\   
\botrule
\end{tabular}
\end{table}

The following example illustrates the usage of \code{hclust\_hurst} function and its \code{print} and \code{plot} methods for multifractional processes. First, we simulate multifractional realizations using the three different Hurst functions defined earlier (five realizations for each).
\begin{lstlisting}
X.list.1 <- replicate(5, GHBMP(t3, H1), simplify = FALSE) #1-5
X.list.2 <- replicate(5, GHBMP(t3, H2), simplify = FALSE) #6-10
X.list.3 <- replicate(5, GHBMP(t3, H3), simplify = FALSE) #11-15
X.list <- c(X.list.1, X.list.2, X.list.3)  
\end{lstlisting}

\begin{figure}[hbt!]
\centering
\includegraphics[width=1\textwidth,height=7cm]{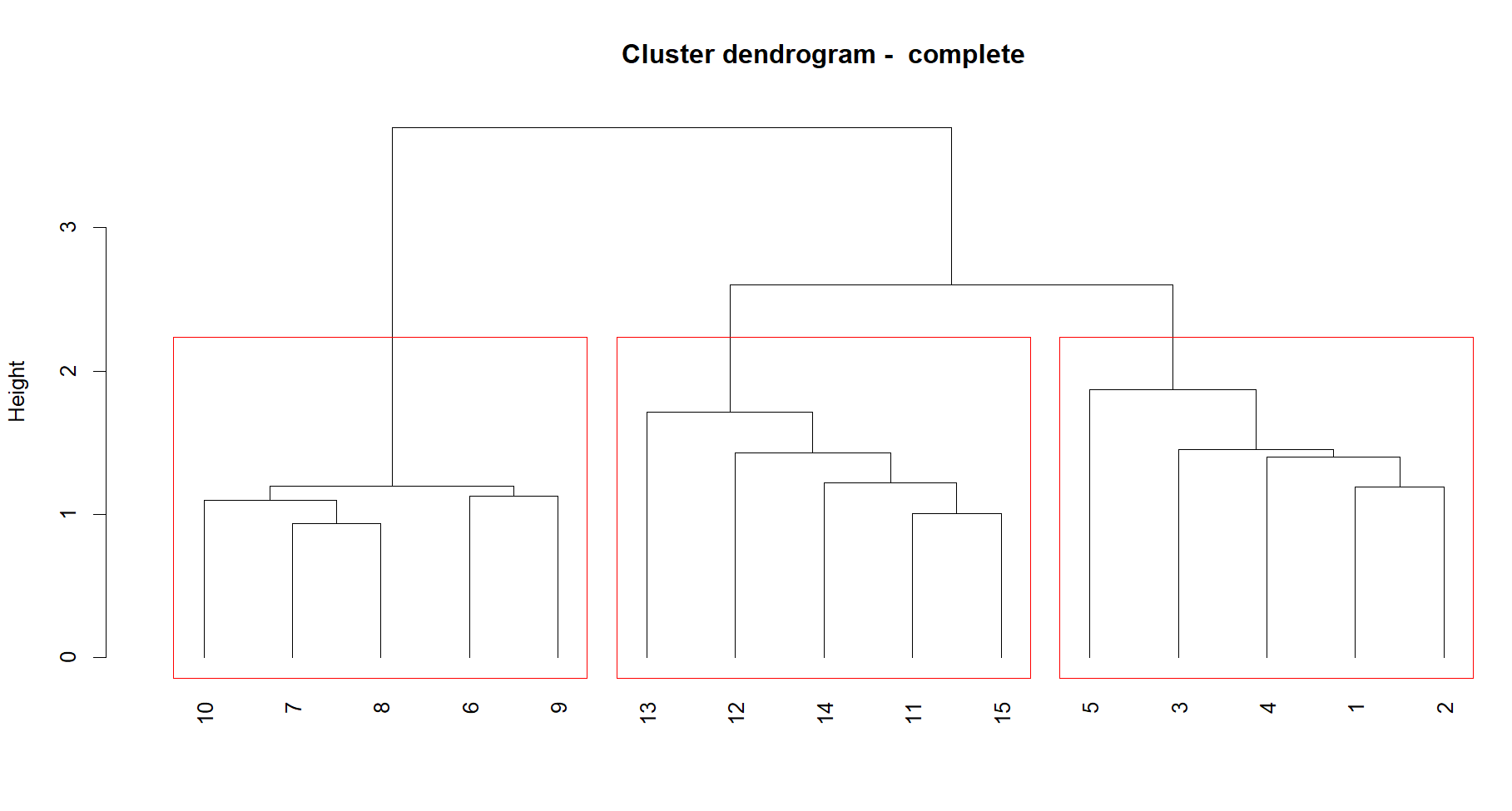}
\caption{\label{fig5} Hierarchical clustering dendrogram of realizations}
\end{figure}
Then \code{hclust\_hurst}  is called to cluster the realizations into three clusters with the dendrogram plotted (see Figure~\ref{fig5}).

\begin{lstlisting}
HC <- hclust_hurst(X.list, k = 3, dendrogram = TRUE)
print(HC)  
Hierarchical clustering with 3 clusters of sizes 5, 5, 5
Clustering information with the distance from the cluster center:
   Item  Cluster  Distance_from_center
1     1        1             0.7830701
2     2        1             0.9181418
3     3        1             0.9106251
4     4        1             0.9099113
5     5        1             1.1185482
6     6        2             0.7529223
7     7        2             0.6438495
8     8        2             0.6674412
9     9        2             0.7220118
10    10       2             0.7314154
11    11       3             0.6684893
12    12       3             0.8913102
13    13       3             1.1275762
14    14       3             0.8856299
15    15       3             0.7074372
Other available components:
[1] "cluster_info" "cluster"
[3] "cluster_sizes" "centers"
[5] "smoothed_Hurst_estimates" "raw_Hurst_estimates"
[7] "call"
\end{lstlisting}

\begin{figure}[hbt!]
\centering
\includegraphics[width=1\textwidth,height=8cm]{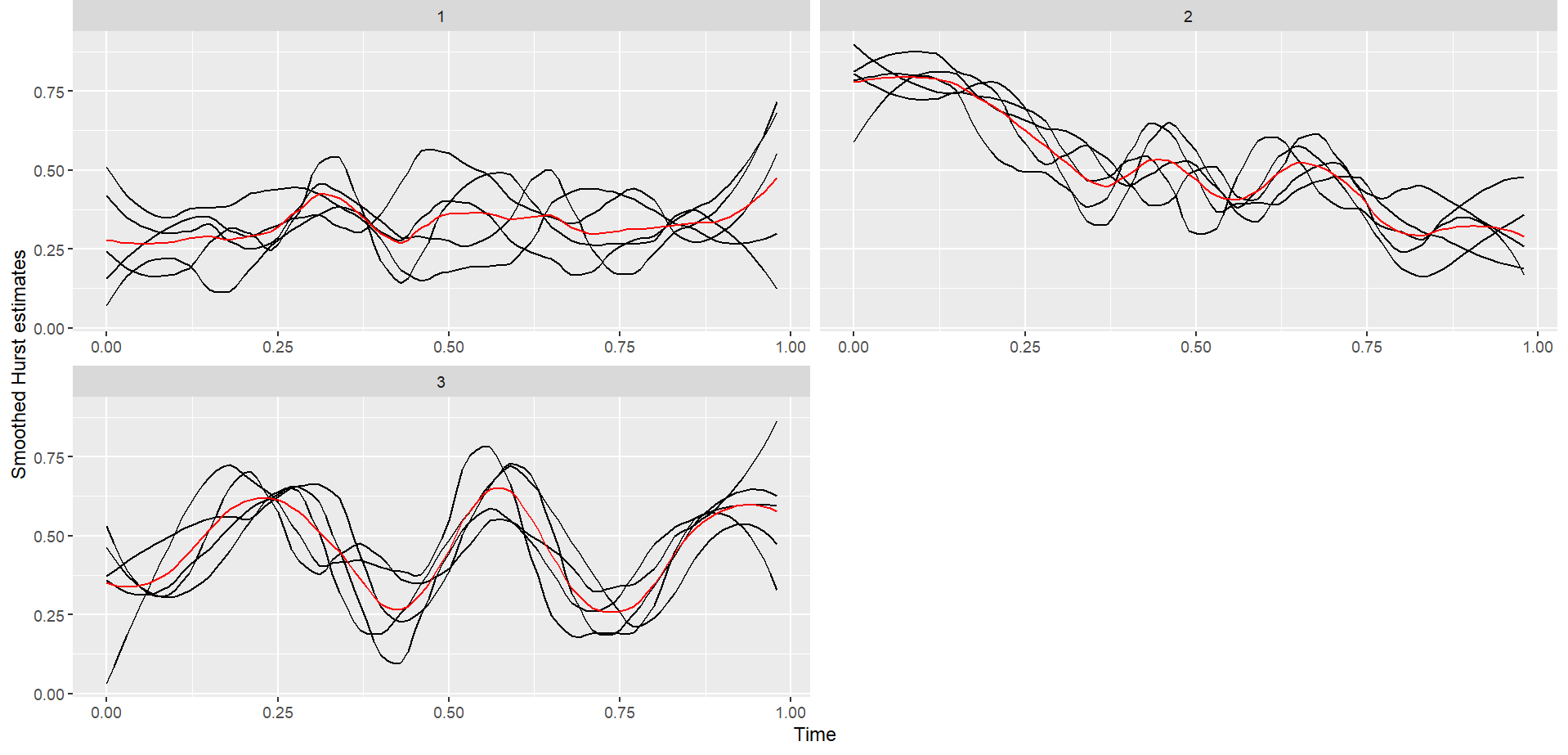}
\caption{\label{fig6} Smoothed Hurst functions of realizations in each cluster and cluster centers}
\end{figure}

Finally, to plot smoothed Hurst functions in each cluster with the corresponding cluster centers in red (see Figure~\ref{fig6}), use the following code 
\begin{lstlisting}
plot(HC, type = "ec")
\end{lstlisting}

The \code{kmeans\_hurst} function implements k-means clustering in the package. The usage is \code{kmeans\_hurst(X.t, k, ..., N = 100, Q = 2, L = 2)}. For a detailed description of the arguments, refer to Table~\ref{table3}. The unspecified argument, denoted as \code{...}, refer to the optional arguments of the \code{kmeans} function in the \pkg{stats} base package, namely, \code{iter.max}, \code{nstart} and \code{algorithm} \citep{Rmanual2025}. The \code{iter.max} argument refers to the number of maximum iterations allowed for one run of the algorithm, and the default is set to $10$. The argument \code{nstart} refers to the number of times the algorithm runs with different random initial centers and the default is $1$. The options available for \code{algorithm} are \code{"Hartigan-Wong"} (default), \code{"Lloyd"}, \code{"Forgy"}, and \code{"MacQueen"}. Refer to the \code{kmeans} function documentation for further details regarding the algorithms. The return is an object of class \code{"k\_hurst"}, which is a list with the same components as in \code{hclust\_hurst}. The return includes the same \code{print} and \code{plot} methods as in \code{hclust\_hurst}.

Finally, we illustrate three functions from the group \textit{Additional functions} to compute geometric statistics of multifractional data. Notice that the functions in this group can be applied to the realization of any random process or time series. 

The function \code{sojourn(X, A, N = 10000, level = "greater", subI = NULL, plot = FALSE)} can be used to estimate the sojourn measure for a realization of a random process or a time series over or under (use "greater" or "lower" respectively for the argument level) a certain constant level \code{A}. The realization \code{X} needs to be provided as a data frame where the first column is a time sequence and the second column includes the sampled values of \code{X}. Providing a vector with upper and lower bounds for \code{subI} one can compute the estimates of sojourn measures for sub-intervals. The argument \code{N} is the number of sub-intervals the time interval is to be split into for computations. Then, if the value of a realization is not available in the corresponding points, linear interpolation is used. After that, to estimate the sojourn measure formula (\rm {\ref{eq3}}) is applied. If the \code{plot} logical option is \code{TRUE}, then the realization, constant level and the excursion region (on the horizontal axis) are plotted.

\begin{figure}[hbt!]
  \centering

  \begin{subfigure}[b]{0.48\textwidth}
    \includegraphics[width=\textwidth,height=6cm]{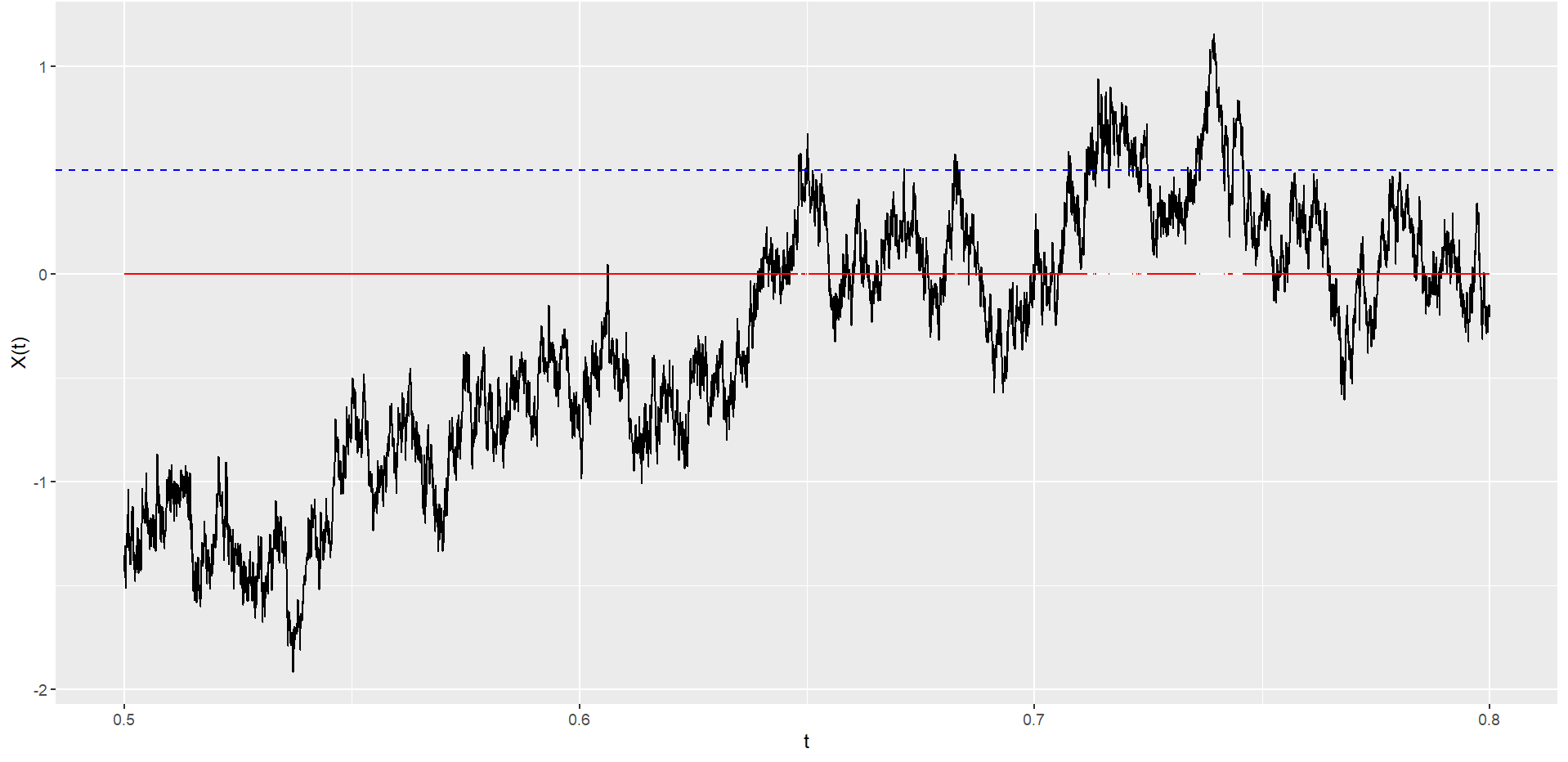}
    \label{fig71}
  \end{subfigure}
   \hspace{0.01\textwidth}
  \begin{subfigure}[b]{0.48\textwidth}
    \includegraphics[width=\textwidth,height=6cm]{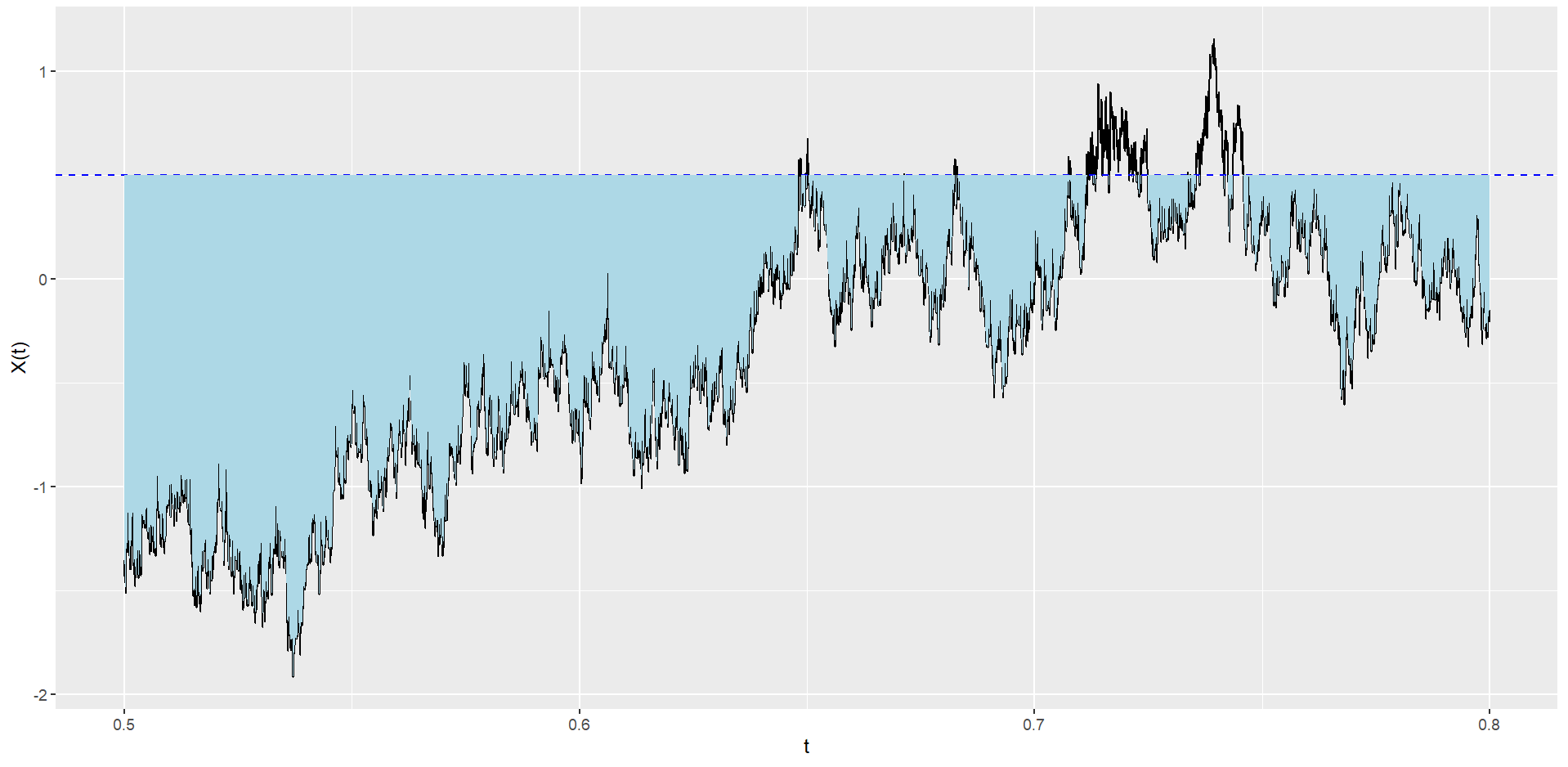}
    \label{fig72}
  \end{subfigure}
  \caption{Excursion region (left) and excursion area (right)
 under the level of $0.5$}
  \label{fig7}
\end{figure}

The function \code{exc\_Area} computes an estimate for the excursion area of a realization over or under (arguments \code{"greater"} or \code{"lower"} respectively) a certain constant level \code{A} by using (\rm {\ref{eq4}}). All the arguments for \code{exc\_Area} are similar to \code{sojourn}. The following code illustrates the usage of these two functions using \code{Process\_1} and Figure~\ref{fig7} shows the corresponding plots.  

\begin{lstlisting}
sojourn(Process_1, A = 0.5, level = `lower', subI = c(0.5, 0.8), plot = TRUE)
[1] 0.2774805
\end{lstlisting}

\begin{lstlisting}
exc_Area(Process_1, A = 0.5, level = `lower', subI = c(0.5,0.8), plot = TRUE)
[1] 0.247952
\end{lstlisting}

\begin{figure}[t!]
\centering
\includegraphics[width=1\textwidth]{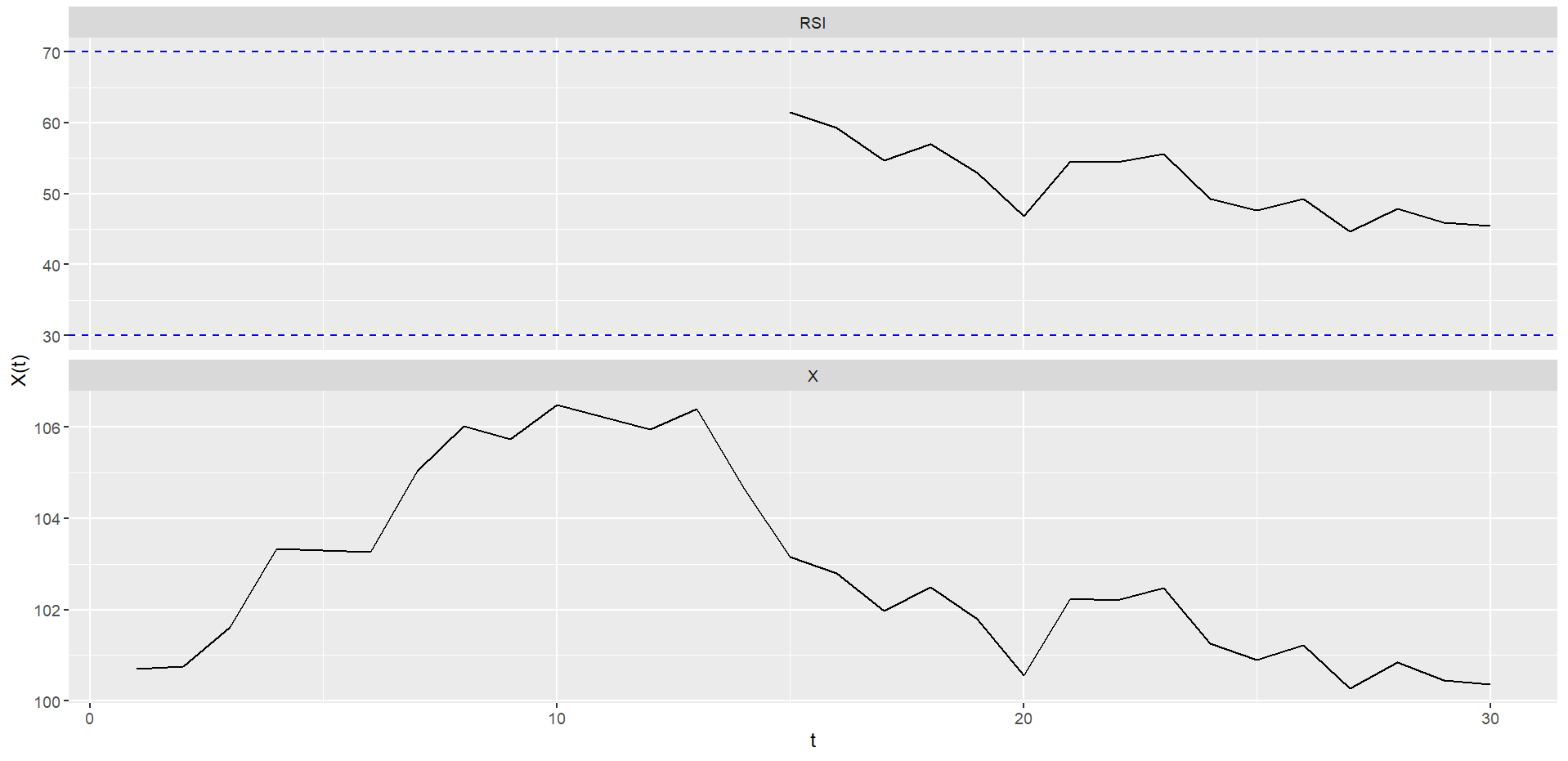}
\caption{\label{fig8} RSI with overbought/oversold thresholds (top) and corresponding price series (bottom)}
\end{figure}

To illustrate the usage of the \code{RS\_Index} function for computing the Relative Strength Index (RSI), we present an example using daily closing prices of a stock. By applying the function with a 14–day look-back period, the output is a vector of RSI values corresponding to the input price time series. Additionally, the function offers an option to generate a two–panel plot as in Figure~\ref{fig8}, with the upper panel displaying the RSI curve with the conventional overbought and oversold thresholds, and the lower panel showing the corresponding price time series.

\begin{lstlisting}
prices <- c(100.70, 100.76, 101.61, 103.33, 103.30, 103.26, 105.04, 106.01, 105.74, 106.48, 106.22, 105.95, 106.39, 104.68, 103.16, 102.79, 101.98, 102.49, 101.79, 100.57, 102.24, 102.21, 102.48, 101.26, 100.91, 101.22, 100.27, 100.85, 100.45, 100.36)
RS_Index(prices, period = 14, plot = TRUE, overbought = 70, oversold = 30)
[1]       NA       NA       NA       NA       NA       NA       
[7]       NA       NA       NA       NA       NA       NA  
[13]      NA       NA   61.53846 59.32109 54.67637 56.96133
[19]  53.01101 46.90549 54.61171 54.45880 55.66208 49.32085
[25]  47.64392 49.28855 44.65883 47.87783 45.89514 45.43918
\end{lstlisting}

The first few values are \code{NA}, since the moving averages required for the calculation cannot be computed until the specified period is reached. The subsequent values capture the relative momentum of the price series: higher RSI values (above 70) indicate potential overbought conditions, while lower ones (below 30) suggest a potential oversold situation.

\subsection{Shiny app for simulation and analysis of GHBMP} \label{sec3.3}

\begin{figure}[htb!]
\centering
\includegraphics[width=1\textwidth]{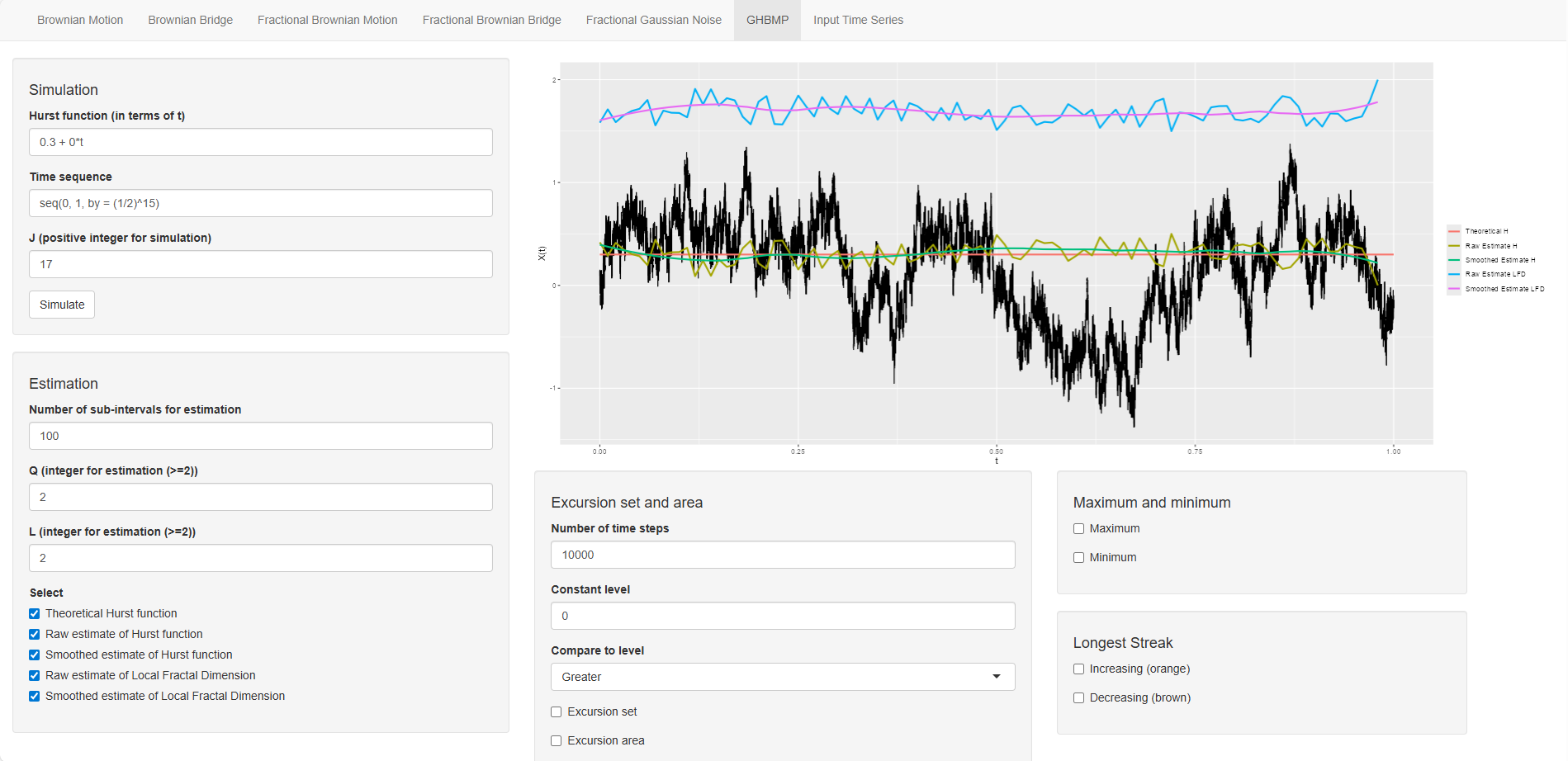}
\caption{\label{fig9} Shiny application interface}
\end{figure}

The \pkg{Rmfrac} package includes an interactive Shiny app designed to use some package functionality via Graphical User Interface (GUI). This includes tabs for simulation and analyses of Bm, Brownian bridge, fBm, fractional Brownian bridge, fractional Gaussian noise, Gaussian Haar-based multifractional processes, and a tab for visualization and analysis of a user-provided time series. In the last case, the time series data need to be uploaded as a CSV file with headers, which includes time and time series values in the first and second columns, respectively. The app provides options to change the simulation parameters and explore changes in the roughness of sample paths in the~GUI. Analysis includes visualization of estimated Hurst functions and fractal dimensions with various geometric statistics. To obtain reliable estimation results for Hurst and fractal dimensions, simulate or upload data with at least 500 data points. In the simulation panel, changes to input values are applied when the \code{Submit} button is clicked, preventing automatic update with every input change. Changes in the analysis panels automatically trigger any simulation (data upload) update and visualize the updated Hurst functions. To launch the Shiny app, call \code{shinyapp\_sim()}. Figure~\ref{fig9} provides a screen capture of the interface of the app. 

\section{Conclusion}\label{sec4}

The paper presents the R package \pkg{Rmfrac} for simulation and analysis of multifractional processes and time series. It provides a comprehensive set of functions for simulating standard and fractional Brownian motions, their bridges, and multifractional processes for a broad range of Hurst functions. Statistical functions include estimation of time-varying roughness, analysis of non-stationary covariance structures, and clustering of realizations based on their roughness. Several other geometric statistics for time series realizations are also available. The package offers various visualization options, including a Shiny application, to explore computational results for simulated and user-provided time series.

Some planned future theoretical research may further expand \pkg{Rmfrac}'s modeling capabilities and practical applications. Generalizing the developed approach to multidimensional cases could enable similar simulations and analysis for random fields and spatial data, see \cite{BROADBRIDGE}. Other series representations, see, for example, \cite{ayache2007wavelet} and \cite{ayache2025harmonizable}, for multi-fractional univariate and multidimensional processes, could be used in the development of new simulation functions. New statistical functions may be based on methodological developments in constructing confidence intervals and statistical tests for Hurst function estimates. 

\backmatter

\bmhead{Acknowledgements}

This research was partially supported by the Australian Research Council's Discovery Projects funding scheme (project number DP220101680). A.Olenko is grateful to Laboratoire d'Excel\-lence, Centre Europ\'{e}en pour les Math\'{e}matiques, la Physique et leurs interactions (CEMPI, ANR-11-LABX-0007-01), Laboratoire de Math\'{e}matiques Paul Painlev\'{e}, France, for support and the opportunity to conduct research at the Universit\'{e} de Lille for a month. He was also partially supported by La Trobe University's SCEMS CaRE and Beyond grant. The authors also thank Prof. A.Ayache for insightful discussions on current trends in the theory of multifractional processes.
The authors are grateful to the anonymous referees for their helpful suggestions, which improved several aspects of the presentation of the paper and the package.

Some performance tests of the package functions were conducted using the Linux computational cluster Gadi of the National Computational Infrastructure (NCI), Australia.

\section*{Declarations}


\begin{itemize}
\item Funding: Australian Research Council's Discovery Projects funding scheme (project number DP220101680).
\item Competing interests: The authors declare no competing interests.
\item Ethics approval and consent to participate: Not applicable.
\item Consent for publication: Not applicable.
\item Data availability: Not applicable.
\item Materials availability: Not applicable.
\item Code availability: The numeric results were obtained using the software R version~4.5.0. The R code is freely available in the folder ``Research materials'' from the website \url{https://sites.google.com/site/olenkoandriy/}.
\item Author contribution: All authors contributed equally to this work.
\end{itemize}








\bibliography{References}

\end{document}